\documentclass[aps,prx,reprint,superscriptaddress,twocolumn,nofootinbib]{revtex4-2}
\usepackage[dvipsnames]{xcolor}
\usepackage{blindtext}
\usepackage{lipsum}
\usepackage{graphics}
\usepackage{amsmath}
\usepackage{graphicx}
\usepackage{graphics}
\usepackage{amssymb}
\usepackage{verbatim}
\usepackage{framed}
\usepackage{float}
\usepackage{braket}
\usepackage{bm}
\usepackage{mathtools}
\usepackage{bbm}
\usepackage{multirow}
\usepackage[toc,page]{appendix}
\usepackage[english]{babel}
\usepackage{CJKutf8}

\usepackage[colorlinks=true,citecolor=dodgerblue,linkcolor=dodgerblue,urlcolor=dodgerblue,pdftitle={AnyonSC}]{hyperref}

\usepackage{tikz} 
\usetikzlibrary{arrows,decorations.pathmorphing,backgrounds,positioning,fit,matrix,calc}
\tikzset{snake it/.style={decorate, decoration=snake,segment length=0.1cm}}

\makeatletter
\def\l@subsubsection#1#2{} 
\makeatother

\usepackage[normalem]{ulem}
\usepackage{wasysym} 
\usepackage[export]{adjustbox}

\makeatletter
\DeclareFontFamily{OMX}{MnSymbolE}{}
\DeclareSymbolFont{MnLargeSymbols}{OMX}{MnSymbolE}{m}{n}
\SetSymbolFont{MnLargeSymbols}{bold}{OMX}{MnSymbolE}{b}{n}
\DeclareFontShape{OMX}{MnSymbolE}{m}{n}{
    <-6>  MnSymbolE5
   <6-7>  MnSymbolE6
   <7-8>  MnSymbolE7
   <8-9>  MnSymbolE8
   <9-10> MnSymbolE9
  <10-12> MnSymbolE10
  <12->   MnSymbolE12
}{}
\DeclareFontShape{OMX}{MnSymbolE}{b}{n}{
    <-6>  MnSymbolE-Bold5
   <6-7>  MnSymbolE-Bold6
   <7-8>  MnSymbolE-Bold7
   <8-9>  MnSymbolE-Bold8
   <9-10> MnSymbolE-Bold9
  <10-12> MnSymbolE-Bold10
  <12->   MnSymbolE-Bold12
}{}

\let\llangle\@undefined
\let\rrangle\@undefined
\DeclareMathDelimiter{\llangle}{\mathopen}%
                     {MnLargeSymbols}{'164}{MnLargeSymbols}{'164}
\DeclareMathDelimiter{\rrangle}{\mathclose}%
                     {MnLargeSymbols}{'171}{MnLargeSymbols}{'171}
\makeatother

\usepackage{ulem}

\usepackage{amssymb}
\usepackage{pifont}

\newcommand{\bmr}{\bm{r}}
\newcommand{\bma}{\bm{a}}
\newcommand{\bmb}{\bm{b}}

\newcommand{\mcD}{\mathcal{D}}

\newcommand{\mcL}{\mathcal{L}}
\newcommand{\mcO}{\mathcal{O}}
\newcommand{\mcP}{\mathcal{P}}

\newcommand{\mcZ}{\mathcal{Z}}

\definecolor{orange(ryb)}{HTML}{FFA500}
\definecolor{dodgerblue}{HTML}{1E90FF}
\definecolor{pinkerton}{HTML}{EC368D}
\definecolor{forest}{HTML}{6DD189}

\usepackage[normalem]{ulem} 

\definecolor{edit}{HTML}{000000}

\usepackage{tikz}
\usepackage{tikz-cd}
\usetikzlibrary{arrows}
\usetikzlibrary{snakes}
\usetikzlibrary{intersections}
\usetikzlibrary{shapes.geometric}
\usetikzlibrary{decorations.pathmorphing, patterns,shapes}
\usetikzlibrary{decorations.markings}

\tikzset{
	partial ellipse/.style args={#1:#2:#3}{
		insert path={+ (#1:#3) arc (#1:#2:#3)}
	}
}

\tikzset{
	mid arrow/.style={postaction={decorate,decoration={
				markings,
				mark=at position .575 with {\arrow[#1]{stealth}}
	}}},
	near arrow/.style={postaction={decorate,decoration={
				markings,
				mark=at position .275 with {\arrow[#1]{stealth}}
	}}},
	far arrow/.style={postaction={decorate,decoration={
				markings,
				mark=at position .800 with {\arrow[#1]{stealth}}
	}}},
}

\pgfdeclarepatternformonly{south west lines}{\pgfqpoint{-0pt}{-0pt}}{\pgfqpoint{3pt}{3pt}}{\pgfqpoint{3pt}{3pt}}{
	\pgfsetlinewidth{0.4pt}
	\pgfpathmoveto{\pgfqpoint{0pt}{0pt}}
	\pgfpathlineto{\pgfqpoint{3pt}{3pt}}
	\pgfpathmoveto{\pgfqpoint{2.8pt}{-.2pt}}
	\pgfpathlineto{\pgfqpoint{3.2pt}{.2pt}}
	\pgfpathmoveto{\pgfqpoint{-.2pt}{2.8pt}}
	\pgfpathlineto{\pgfqpoint{.2pt}{3.2pt}}
	\pgfusepath{stroke}}

\makeatletter
\renewcommand\onecolumngrid{
\do@columngrid{one}{\@ne}%
\def\set@footnotewidth{\onecolumngrid}
\def\footnoterule{\kern-6pt\hrule width 1.5in\kern6pt}%
}

\renewcommand\twocolumngrid{
        \def\footnoterule{
        \dimen@\skip\footins\divide\dimen@\thr@@
        \kern-\dimen@\hrule width.5in\kern\dimen@}
        \do@columngrid{mlt}{\tw@}
}
\makeatother

\begin{document}

\title{Anyon Superconductivity from Topological Criticality in a Hofstadter-Hubbard Model}

\author{Stefan Divic}
\affiliation{Department of Physics, University of California, Berkeley, CA 94720, USA}

\author{Valentin Cr\'epel}
\affiliation{Center for Computational Quantum Physics, Flatiron Institute, New York, New York 10010, USA}

\author{Tomohiro Soejima (\begin{CJK*}{UTF8}{bsmi}副島智大\end{CJK*})}
\affiliation{Department of Physics, Harvard University, Cambridge, MA 02138, USA}

\author{Xue-Yang Song}
\affiliation{Department of Physics, Hong Kong University of Science and Technology, Clear Water Bay, Hong Kong, China}

\author{Andrew J. Millis}
\affiliation{Center for Computational Quantum Physics, Flatiron Institute, New York, New York 10010, USA}
\affiliation{Department of Physics, Columbia University, New York, NY 10027, USA}

\author{Michael P. Zaletel}
\affiliation{Department of Physics, University of California, Berkeley, CA 94720, USA}
\affiliation{Material Science Division, Lawrence Berkeley National Laboratory, Berkeley, CA 94720, USA}

\author{Ashvin Vishwanath}
\affiliation{Department of Physics, Harvard University, Cambridge, MA 02138, USA}

\date{\today}

\begin{abstract}
We argue that the combination of strong repulsive interactions and high magnetic fields can generate electron pairing and superconductivity. Inspired by the large lattice constants of moir\'e materials, which make large flux per unit cell accessible at laboratory fields, we study the triangular lattice Hofstadter-Hubbard model at one-quarter flux quantum per plaquette, where previous literature has argued that a chiral spin liquid separates a weak-coupling integer quantum Hall phase and a strong-coupling topologically-trivial antiferromagnetic insulator at a density of one electron per site. We argue that topological superconductivity emerges upon doping in the vicinity of the integer quantum Hall to chiral spin liquid transition. We employ exact diagonalization and density matrix renormalization group methods to examine this theoretical scenario and find that electronic pairing indeed occurs on both sides of criticality over a remarkably broad range of interaction strengths. On the chiral spin liquid side, our results provide a concrete model realization of the long-hypothesized mechanism of anyon superconductivity. Our study thus establishes a beyond-BCS mechanism for electron pairing in a well-controlled limit, relying crucially on the interplay between electron correlations and band topology.
\end{abstract}

\maketitle

\newcommand{\nocontentsline}[3]{}
\begingroup
\let\addcontentsline=\nocontentsline

\section{Introduction}

The search for electron pairing mechanisms that go beyond the well-established BCS/``pairing glue'' paradigm, exemplified by the electron-phonon mechanism~\cite{Frohlich1950,Bardeen1950,BCS1957}, has a long history.
A popular theoretical route has been to consider situations where charge is associated with topological excitations~\cite{Skyrme1961, ANDERSON1973, FazekasAnderson1974, SSH1979, Laughlin1983, Kivelson1987, LeeKane1990, Sondhi1993, AbanovWiegmann2001,GroverSenthil,ChatterjeeBultinckZaletel2020,Khalaf_2021}. The best known of these proposals are the resonating valence bond (RVB)~\cite{Anderson1987,Kivelson1987,BASKARAN1987,KotliarLiu1988} and anyon superconductivity scenarios~\cite{laughlin_superconducting_1988,laughlin_relationship_1988,fetter_random_phase_1989,ChenWilczekWittenHalperin1989,WenWilczekZee1989,WenZee1989,WenZee1990,lee_anyon_1989,HosotaniChakravarty,BanksLykken1990}, where doping a quantum spin liquid with fractionalized charge excitations leads to superconductivity. In the anyon superconductivity approach proposed soon after the discovery of high-$T_c$ superconductivity in the layered cuprates, the starting point is the chiral spin liquid (CSL) phase introduced by Kalmeyer and Laughlin~\cite{Kalmeyer87}.
Although interest in the anyon superconductivity mechanism in the context of high-$T_c$ cuprates has waned,
it remains a remarkable theoretical example of superconductivity emerging from a chiral insulator.

Spin liquids have proven elusive, and even where they are proposed to appear, superconductivity need not arise upon doping. The triangular lattice Hubbard model with time reversal symmetry has been reported to host a CSL ground state at intermediate coupling~\cite{Szasz2020,Szasz2021,Chen2022,Kadow2022}. However, various magnetic orders and other spin liquid phases are so close in energy to the CSL~\cite{Gong2017,WietekHeisenbergChiral2017,SaadatmandMcCulloch2017,Cookmeyer2021} that its existence in this model, and in related time reversal-symmetric extended Heisenberg models~\cite{Cookmeyer2021,SaadatmandMcCulloch2017,ZhuWhite2015,Gong2019,JiangJiang_2023}, is not definitively established~\cite{Shirakawa2017,Wietek2021}. Further, the existence of superconductivity at small hole doping in the intermediate-coupling regime~\cite{Chen2013,Zampronio2023} relevant to the putative CSL has not been clearly demonstrated. Indeed, a density matrix renormalization group (DMRG) analysis reports a metal rather than a superconductor~\cite{ZhuShengVishwanath2022}. In addition, there is no spin liquid in the weak-coupling regime where superconductivity has also been proposed~\cite{Raghu2010,Gannot2020,ArovasHubbardReview2022}.

In this work, we show that superconductivity emerges naturally in the vicinity of \textit{topological criticality} arising from a continuous transition between two topologically-distinct insulators, with time reversal symmetry explicitly broken by a magnetic field. The crucial point is that a change in topology requires closing and reopening a gap at the transition, in this case a charge gap associated with a bosonic charge-$2e$ mode. Near the topological critical point, these bosonic modes are the lowest-energy local charge excitations, while unpaired electronic states appear only at higher energies. Modest doping then introduces paired carriers, which can superconduct.

We illustrate the potential of this idea through explicit calculations in a microscopic Hofstadter-Hubbard model argued to host an integer quantum Hall (IQH) phase at weak coupling with charge and spin edge modes, and a CSL phase with only a spin edge mode, arising upon increasing electron correlations via repulsive Hubbard interactions~\cite{Kuhlenkamp2024,Divic2024}. The CSL phase is also numerically observed in extended Heisenberg models with a chiral spin interaction~\cite{Gong2017,SaadatmandMcCulloch2017,WietekHeisenbergChiral2017} arising from the orbital magnetic flux through each plaquette~\cite{SenChitra1995,Motrunich2006}. The IQH and CSL insulators appear at a density of one electron per site, henceforth referred to as ``half filling,'' with superconductivity emerging upon doping electrons or holes. In Fig.~\ref{fig:main_ideas}(a), we present a schematic phase diagram in the plane of chemical potential $\mu$ and Hubbard $U$.

Our study goes beyond previous proposals for superconductivity in the Hofstadter flux regime, which considered only attractive~\cite{maska_reentrant_2002, mo_fermionpairing_2002, zhai_pairing_2010, iskin_stripeordered_2015, jeon_topological_2019, sohal_intertwined_2020, schirmer_phase_2022} and perturbatively weak repulsive interactions~\cite{shaffer_unconventional_2022}. The investigation of the present strongly-repulsive model is motivated by its potential realization in transition metal dichalcogenide (TMD) moir\'e materials~\cite{Zhang2021,Kuhlenkamp2024}, whose advent has opened new avenues for realizing exotic quantum phases and enabled the controlled variation of their doping~\cite{Balents_review2020,MakShan_review2022,NuckollsYazdani_review2024}. The large unit cells of these materials make the Hofstadter flux regime experimentally accessible, while the Zeeman coupling can be quenched~\cite{Kuhlenkamp2024}.

We present general arguments supported by a parton treatment and effective field theory that reveal low-energy Cooper pairs near the putative IQH-CSL critical point. Furthermore, we contend that doping in its vicinity leads to superconductivity. On the CSL side, the superconductor is shown to arise via the \textit{anyon superconductivity} mechanism. Pairing is then examined numerically with exact diagonalization (ED) and DMRG, which reveal that charge-$2e$ excitations are indeed the lowest-energy local charge excitations, per unit charge, over a broad range of parameters, a compelling precursor to superconductivity upon doping. Strikingly, the pairing extends beyond the CSL, well into the IQH side of the phase diagram where anyons cannot be invoked, consistent with the softening of the $2e$ gap by topological criticality.

Importantly, this mechanism does not rely on strict quantum criticality. Superconductivity emerges in a non-infinitesimal neighborhood of the topological critical point and will therefore persist even if the transition is weakly-first-order or one of the phases is proximal but avoided. Its precursor, electron pairing, appears even in the finite systems accessible in our numerics. Altogether, our results suggest that superconductivity can arise in a regime with both strong repulsive interactions and broken time reversal symmetry, a combination that normally disfavors superconductivity.

The remainder of the manuscript is organized as follows. In Sec.~\ref{sec:model_symmetries}, we introduce the Hofstadter-Hubbard model as well as its continuous and magnetic space group symmetries. In Sec.~\ref{sec:partons}, we provide a parton description of the putative topological critical point and the mechanism for obtaining superconductivity upon doping each side of the transition. In Sec.~\ref{sec:numerical_evidence}, we provide numerical evidence for electron pairing using both ED and DMRG. Sec.~\ref{sec:Discussion} is a summary and conclusion suggesting avenues for future research. The \textit{SI Appendix} provides details of calculations described in the main text and supporting data.

\begin{figure}
\centering
\includegraphics[width=\linewidth]{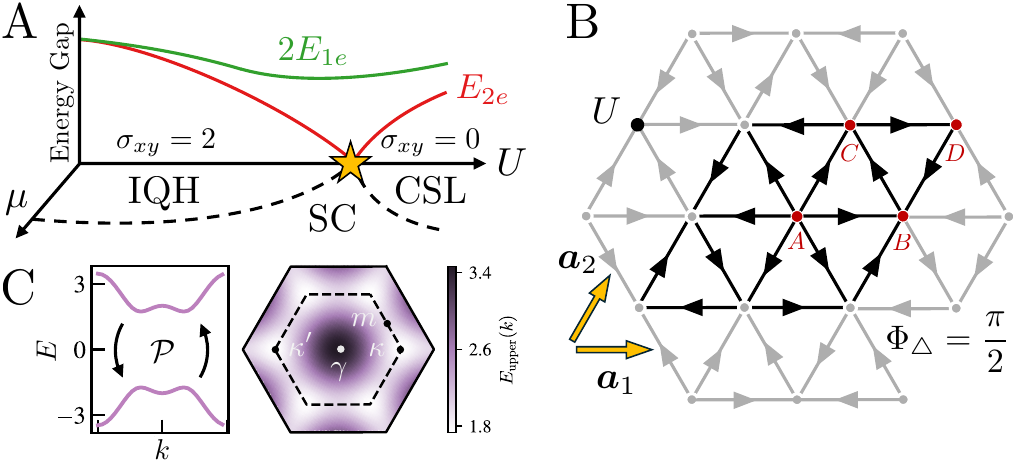}
\caption{
    (a) Schematic depiction of the predicted energy gap to $2e$ excitations (red) and twice the energy of $1e$ excitations (green) at half filling of the Hofstadter-Hubbard model with $\Phi_\triangle=\pi/2$ flux per triangle as a function of interaction strength $U$ in the 2D limit. Proposed phase diagram in the plane of $U$ and chemical potential $\mu$. The star indicates a topological quantum phase transition from an integer quantum Hall (IQH) insulator to chiral spin liquid (CSL) at half filling. The dashed lines show the chemical potential required to add carriers and produce a chiral superconductor (SC) with topologically-protected edge modes.
    (b) Finite region of the two-dimensional model with $\Phi_\triangle=\pi/2$ flux per triangle, repulsive on-site interactions $U$, and Bravais vectors $\bm{a}_{1,2}$ indicated. The hoppings may be chosen to be $C_{6}$-invariant and imaginary (each arrow indicates an amplitude $-it$), with a four-site unit cell (sublattices $A$-$D$ in red).
    (c) Left: band structure at  $\Phi_\triangle=\pi/2$ consisting of two bands related by particle-hole symmetry $\mathcal{P}$. Right: two-dimensional color map of energy as a function of momentum for the upper energy band over the Brillouin zone of the $2\times 2$ unit cell (note the bands are two-fold degenerate at every momentum).
    }
\label{fig:main_ideas}
\end{figure}

\section{Model and Symmetries} \label{sec:model_symmetries}

The Hofstadter-Hubbard model (see Fig.~\ref{fig:main_ideas}b), with hoppings $t_{ij}$ from site $j$ to $i$ and on-site interaction $H_I$, is given by $H_\text{HH} = -\sum_{ij} t_{ij}c^\dag_i c_j + H_I$. The hopping amplitudes $t_{ij}$ are complex numbers with phases encoding the orbital magnetic flux~\cite{Peierls1933}. In this work, we specialize to the triangular lattice with $\Phi_\triangle=\pi/2$ flux per triangle~\cite{Kuhlenkamp2024,Divic2024} for which the hopping amplitudes can be chosen to be purely imaginary, so that the single-particle term in the Hamiltonian may be written as:
\begin{align} \label{eq:hopping}
     H_0 = i\sum_{\langle ij\rangle}\tau_{ij}\left ( c^\dagger_{i\sigma} c_{j\sigma} -c^\dagger_{j\sigma} c_{i\sigma}\right ).
\end{align}
The arrows on bonds in Fig.~\ref{fig:main_ideas}(b) indicate our sign convention that the hopping from site $j$ to $i$ along an arrow has $\tau_{ij} = -t < 0$, while hopping opposite an arrow has the opposite sign. We report energies in units of $t$.

The symmetries of $H_0$ are most clearly exposed by mapping to Majorana fermions, $c_{j\uparrow}=\frac12 (\chi_1^j+i\chi_2^j)$ and $c_{j\downarrow}=\frac12 (\chi_3^j+i\chi_4^j)$, where $\{\chi_a^i,\,\chi_b^j\} = 2\delta_{ab}\delta_{ij}$
so that $H_0$ takes the form~\cite{Affleck1990}:
\begin{align}
    H_0 = \frac{i}{2}\sum_{1\leq a \leq 4} \sum_{\langle ij\rangle} \tau_{ij} \chi^i_a \chi^j_a,
\end{align}
which is seen to have an O(4) symmetry, corresponding to proper/improper rotations of the Majorana four-component vector.
The Hubbard interaction
\begin{align} \label{eq:Hubbard}
H_I = U \sum_{i} (n_{i,\uparrow} -1/2)(n_{i,\downarrow} -1/2)
\end{align}
can be written in the Majorana representation as $H_I = -(U/4) \sum_i \chi^i_1 \chi^i_2 \chi^i_3 \chi^i_4$.
Note that the interaction breaks the O(4) symmetry of the hopping model down to SO(4). The operators in the determinant $-1$ sector, corresponding to improper rotations, are related to particle-hole transformations that flip the sign of $U$ (\textit{e.g.}, conjugating only the spin-up electrons) and are therefore not symmetries of the interacting theory.
The particle-hole operation which \textit{preserves} the sign of $U$ is a symmetry, identified in Ref.~\cite{Divic2024}.

Contained in SO(4) are the spin $\mathrm{SU(2)}_s$ and pseudospin $\mathrm{SU(2)}_c$ symmetry subgroups identified by Yang and Zhang~\cite{Yang1989,1990YangZhang,Yang1991}, and by Affleck~\cite{Affleck1990}. While the pseudospin symmetry is typically associated with bipartite lattice Hubbard models~\cite{Hermele2007}, we highlight that it is more generally present whenever there exists a gauge in which the hoppings are purely imaginary~\cite{WenBook2007}, and that $\mathrm{SU(2)}_c$ reduces to the usual charge $\mathrm{U(1)}_c$ for generic long-range interactions. The $\mathrm{SU(2)}_c$ pseudospin symmetry will play an important role in our study of the topological phase transition in Sec.~\ref{sec:criticality_half_filling}.

The model is additionally symmetric under operations in the magnetic space group generated by the rotation $C_6$ and translation $T_1$ along the nearest-neighbor direction $\bm{a}_1$ (see Fig.~\ref{fig:main_ideas}b). Their explicit form, which we provide in \textit{SI Appendix}, Sec.~1B, depends on the choice of gauge for the electron operators. However, they satisfy certain gauge-invariant generating relations which will prove important for interpreting the pairing symmetry in Sec.~\ref{subsec:pairing_sym}.
Because the magnetic space group is a U$(1)_c$ extension~\cite{FLOREK199481}, the overall U$(1)_c$ phase of each spatial symmetry is ambiguous. We make the following convenient choices: we fix $(C_6)^6 = 1$, choose $T_1$ to satisfy $C_2 T_1 C_2^\dagger = T_1^\dagger$ which constrains it modulo a sign,
and define rotated nearest-neighbor translations by $T_{j+1} = C_6^{j} T_1 C_6^{-j}$.
Moreover, due to the $\pi/2$ flux per triangle, one may verify the following gauge-invariant relations:
\begin{align} \label{eq:mag_algebra_relations}
T_2 T_1 &= (-1)^{N_F} T_1 T_2, \quad T_5 T_3 T_1 = (\pm i)^{N_F},
\end{align}
where $N_F$ is the fermion number. The sign ambiguity in the second relation arises because our conventions do not fix the sign of $T_1$; we remedy this by choosing $+i$.
When acting on charge-$2e$ \textit{pairs}, with $N_F=2$, the first relation is trivial but the second is not. This is the origin of the unorthodox pairing symmetry we uncover in Sec.~\ref{subsec:pairing_sym}.

\section{Parton Theory} 
\label{sec:partons}

\subsection{Topological Criticality at Half Filling} \label{sec:criticality_half_filling}

We introduce a parton construction that captures the effects of on-site $U$, in effect performing a Gutzwiller projection on the IQH wavefunction to obtain the CSL~\cite{He2011,ChenHazzardReyHermele2016}. This formalism reproduces the phase diagram of the half-filled model obtained numerically in Refs.~\cite{Kuhlenkamp2024,Divic2024} and allows us to study the effects of doping. While it is possible to proceed by retaining the $\mathrm{SU(2)}_c$ pseudospin symmetry~\cite{Hermele2007}, we relegate this treatment to \textit{SI Appendix}, Sec.~6 and, for ease of presentation, derive here a slave-rotor theory retaining only $\mathrm{U(1)}_c$ along with $\mathrm{SU(2)}_s$.
We begin by introducing a U(1) rotor variable $e^{i\theta}$ and its conjugate integer-valued ``angular momentum'' $L$ at each site~\cite{FlorensGeorges2002}. We write the electron operator as $c^\dagger_{i,\sigma} = f^\dagger_{i,\sigma} \,e^{i\theta_i}$ where $f$ is a fermionic ``spinon'' carrying the electronic spin. The redundancy introduced by the rotors
is alleviated by imposing the following constraint at each site:
\begin{align} \label{eq:U(1)_constraint}
    L_i = f^\dagger_{i,\uparrow} f_{i,\uparrow}+ f^\dagger_{i,\downarrow} f_{i,\downarrow} -1.
\end{align}
Note that a singly-filled site is represented by $L_i=0$, while the doublon/empty sites correspond to $L_i=\pm 1$. The Hubbard model can then be rewritten as~\cite{FlorensGeorges2002}:
\begin{align} \label{eq:rotor}
H = -\sum_{\langle ij\rangle,\sigma} t_{ij}e^{i(\theta_i-\theta_j)}f^\dagger_{i,\sigma} f_{j,\sigma} +{\rm h.c.} +\frac{U}{2}\sum_{i}L_i^2.
\end{align}
To make progress, we adopt a mean field approach, replacing operator bilinears by their mean field values:
\begin{equation}
\begin{aligned} \label{eq:MF}
H_\mathrm{MF} &= H_f + H_\theta \\
H_{f} &= -\sum_{\langle ij\rangle,\sigma}   \tilde{t}_{ij}  f^\dagger_{i,\sigma} f_{j,\sigma} +{\rm h.c.} \\
H_{\theta} &= -\sum_{\langle ij\rangle,\sigma} {Q}_{ij} e^{i(\theta_i-\theta_j)} + {\rm h.c.} + \frac{U}{2}\sum_{i}L_i^2,
\end{aligned}
\end{equation}
where $Q_{ij} = t_{ij} \langle f^\dagger_{i,\sigma} f_{j,\sigma}\rangle$ and $ \tilde{t}_{ij}  = t_{ij}\langle e^{i(\theta_i-\theta_j)}\rangle $.
Let us summarize the main results, leaving the detailed analysis of this mean field theory to \textit{SI Appendix}, Sec.~2. Consistency with the numerical results of Ref.~\cite{Divic2024}, namely a spin gap throughout the transition, requires that the spinons $f$ be gapped. The most natural possibility, consistent with the projective construction of the CSL in the Mott limit~\cite{WenBook2007}, is that the spinons see $\Phi_\triangle = \pi/2$ net flux per plaquette, like the microscopic electrons, and enter a spin-singlet Chern insulator with total Chern number $C=2$.
Since the rotor bosons then see no net flux,\footnote{This contrasts with the scenario of electrons at zero magnetic field~\cite{song2021doping,SongZhang2023}, where the two parton species experience opposite, non-zero net flux.}
$H_\theta$ displays a superfluid-Mott transition on increasing $U$, corresponding to the electronic IQH-CSL transition. A mean field treatment of $H_\theta$ locates this critical point (see \textit{SI Appendix}, Sec.~2) at $U^*/t \approx 9.6$, close to the transition point $U^*/t \approx 12$ estimated by iDMRG~\cite{Kuhlenkamp2024,Divic2024}.

Using Eq.~\eqref{eq:MF}, we show that the universal response properties of the spinon-rotor mean field solutions are consistent with that of the IQH and CSL. We introduce a gauge field on bonds to account for gauge fluctuations about the mean field solution~\cite{MarstonAffleck1989},
giving ${Q}_{ij} \rightarrow Q e^{-ia_{ij}}$ and $\tilde{t}_{ij} \rightarrow \tilde{t}_{ij} e^{+ia_{ij}}$.
Further, we introduce a gauge field $A$ ($A_s$) coupled minimally to the rotors (spinons) to probe the charge (spin) response. Restricting to energies below the spin gap, we may integrate out the spinons; the unit Chern number for each spin species yields the following Chern-Simons terms~\cite{AlvarezGaumeWitten1984,Redlich1984,qi2008topological}:
\begin{align}
 {\mathcal L}_\text{CSL} = \frac{2}{4\pi} (a\wedge da + A_s\wedge dA_s).
\end{align}
We employ the notation $a\wedge da = \epsilon_{\mu\nu\lambda} a^\mu \partial^\nu a^\lambda$, where summation over Greek spacetime indices is implicit and $\epsilon$ is the three-dimensional antisymmetric tensor.

We now turn to the rotor charge excitations. Since we are interested here in long wavelength properties, it is convenient to pass to the continuum limit and utilize a coarse-grained ``soft-spin'' description~\cite{Fisher1989_bosehubbard}, replacing $\psi \sim e^{i\theta}$ and introducing a potential $V(\psi) = m^2|\psi|^2 + \lambda |\psi|^4$. Note that there is only a single bosonic field $\psi$, corresponding to a single minimum in the rotor dispersion, a consequence of the vanishing average flux experienced by the rotors. Tuning the sign of $m^2$ from positive to negative induces condensation of $\psi$, as we describe below. The resulting effective field theory is:\footnote{While resembling the Zhang-Hansson-Kivelson composite boson description of the FQHE~\cite{ZHK1989,Zhang_1992IJMPB}, Eq.~\eqref{eq:chargon_Lagrangian} differs in its Lorentz-invariance, particle-hole symmetry $(\psi,a)\to (\psi^*,-a)$ absent the probe gauge fields, and integer-level dynamical Chern-Simons term.}
\begin{equation} \label{eq:chargon_Lagrangian}
{\mathcal L} = |{\mathcal D}_{A-a}\psi|^2 - V(\psi) + \frac{2}{4\pi} a\wedge d\, a + \frac{2}{4\pi}A_s\wedge d\, A_s,
\end{equation}
where ${\mathcal D}^\mu_{b} = \partial^\mu - ib^\mu$, and we have taken advantage of field rescaling to bring it into this form. The time component $a_0$ is introduced to implement the on-site constraint Eq.~\eqref{eq:U(1)_constraint}. We now consider the two phases and their transition.

\smallskip
{\bf Phase I:} $m^2<0,\, \langle \psi \rangle \neq 0$. In the small-$U$ limit we expect the rotor variables to ``condense,'' which introduces a Higgs mass term for $a-A$. Integrating over $a$ yields the response theory ${\mathcal L}_{\mathrm{I}}[A, A_s] =\frac{2}{4\pi} A\wedge d A + \frac{2}{4\pi}A_s\wedge d A_s $.
Thus we have an insulator without intrinsic topological order, with precisely the charge and spin quantum Hall conductance of the spin-singlet integer quantum Hall state, namely $\sigma_{xy}=2\cdot e^2/h$ and $\sigma^s_{xy}=2\cdot \hbar/8\pi$. The latter describes the quantized response of spin to a Zeeman gradient, termed the spin quantum Hall effect~\cite{SenthilMarstonFisher1999}, and is distinct from the quantum spin Hall effect~\cite{KaneMele2005,BernevigZhang2006}.

\smallskip
{\bf Phase II:} $m^2>0,\, \langle \psi \rangle =0$. In the larger-$U$ regime, the rotor condensate disappears and the $\psi$ field is gapped. The low-energy effective action in this phase is then ${\mathcal L}_{\mathrm{II}}[a, A, A_s] = \frac{2}{4\pi} a\wedge d a + \frac{2}{4\pi}A_s\wedge d A_s =$ ${\mathcal L}_\text{CSL}$. While the quantized spin response is the same as in Phase I (guaranteed by the spin gap being maintained throughout), the charge Hall conductivity vanishes. Semion topological order now arises from the dynamical U$(1)_2$ Chern-Simons term. These are the characteristic properties of the CSL phase~\cite{WenBook2007}. Moreover, the change of $\sigma_{xy}$ across the IQH-CSL transition implies that the charge gap \textit{must} close if the transition is continuous.

\smallskip
{\bf Critical Point:} In the mean field approximation, the critical point corresponds to setting $m^2=0$ in Eq.~\eqref{eq:chargon_Lagrangian}, so that the $\psi$ field becomes gapless. Intuition for this field is gained by recognizing that the fractionalized spin-$1/2$ semion excitation in the CSL phase combines with the electron to give a semionic charge-$e$ spin-0 excitation, represented by the non-local $\psi$ field. At mean field level, the microscopic particle-hole symmetry ensures that the IQH-CSL transition is continuous and belongs to the 3D XY universality class~\cite{Fisher1989_bosehubbard}.

To characterize the critical point beyond the mean field approximation, and assess whether the transition remains continuous, we consider gauge field fluctuations and the dynamical Chern-Simons term in Eq.~\eqref{eq:chargon_Lagrangian}. This full theory cannot be solved exactly, and the Chern-Simons term introduces a sign problem that would hamper large-scale quantum Monte Carlo simulations~\cite{Golan_intrinsicsign}. Alternatively, one can study the microscopic Hamiltonian of Sec.~\ref{sec:model_symmetries} using tensor network methods. Indeed, the cylinder DMRG studies of Refs.~\cite{Kuhlenkamp2024,Divic2024} provide direct evidence that the transition is continuous and that electron excitations are gapped across the transition. This points to spin-singlet Cooper pairs being the cheapest local charge excitations near the critical point, which we demonstrate numerically in Sec.~\ref{sec:numerical_evidence}.

In addition, we can study deformations of the parton critical theory that \textit{can} be characterized analytically and are known to exhibit a continuous transition. Specifically, consider Eq.~\eqref{eq:chargon_Lagrangian} but with Chern-Simons term $\frac{N}{4\pi}a\wedge da$, while retaining the single bosonic scalar field $\psi$. While we are interested in the case $N=2$, observe that $N = 1$ and $N = 0$ both correspond to continuous transitions: under fermion-boson duality~\cite{DualityWeb2016}, the former describes the free-fermion Dirac critical point separating two insulators with Chern numbers $C=0\to1$, while the latter describes the 2+1D superconductor-insulator transition~\cite{PESKIN1978122,DasguptaHalperin1981}. Furthermore, in the $N=\infty$ limit, the theory reduces to the 3D XY transition since gauge field fluctuations are suppressed. The existence of a continuous transition in the $N=0,1,\infty$ deformed theories suggests that the $N=2$ transition can also be continuous.

We shed further light on the IQH-CSL transition by mapping it to an \textit{equivalent}, better-studied bosonic transition between a trivial insulator and $\nu = -1/2$ Laughlin state of charge-$2e$ Cooper pairs~\cite{WenWu1993,ChenAnyonGas1993,BarkeshliMcGreevy2014,Lee2018}.
The equivalence can be seen by ``stacking'' an invertible $\nu=-2$ IQH phase above Phases I and II~\cite{SongZhang2023}. This trivializes both the spin and charge response of the IQH and maps the CSL to a phase with Hall conductance $\sigma_{xy}=-\frac12 \frac{(2e)^2}{h}$ but no spin response, namely a Laughlin liquid of Cooper pairs. The transition can thus be viewed as a plateau transition of Cooper pairs, rationalizing there being gapless Cooper pairs at the critical point.

This bosonic transition is argued to be continuous, with critical exponents and universal conductivity computed perturbatively, in several works~\cite{WenWu1993,ChenAnyonGas1993,Ye1998}.
In contrast, Ref.~\cite{PryadkoZhang1994} proposes a first-order transition scenario, while acknowledging that a continuous transition remains possible elsewhere on the phase boundary. More recently, DMRG studies of models of hardcore bosons support a continuous transition~\cite{Motruk2017,Zeng2021}, though they do not estimate critical exponents or check for signatures of conformal invariance.

In Ref.~\cite{Lee2018}, the bosonic trivial-to-Laughlin transition theory is expressed both in a form equivalent to Eq.~\eqref{eq:chargon_Lagrangian} and as its fermionic dual, a $\mathrm{QED}_3$-Chern-Simons theory~\cite{BarkeshliMcGreevy2014}. Due to a conjectured ``level-rank'' duality~\cite{Hsin2016JHEP}, the theory is believed to possess an emergent SO(3) symmetry~\cite{Benini2017JHEP,Lee2018}, rotating between the boson density, creation, and annihilation operators, which go gapless at criticality.
Should this critical point be described by a conformal field theory, the putative SO(3) symmetry imposes an important constraint: since the conserved density has protected scaling dimension 2, the Cooper pair insertion operator should have the same scaling dimension~\cite{Lee2018}. Remarkably, our lattice model provides a microscopic realization of this transition in which the pseudospin $\mathrm{SU(2)}_c$ symmetry introduced in Sec.~\ref{sec:model_symmetries} explicitly implements this conjectured SO(3) symmetry. In fact, we exploit the pseudospin symmetry within the non-abelian slave-rotor formalism~\cite{Hermele2007} to derive a bosonic SU$(2)_1$ Chern-Simons theory dual to Eq.~\eqref{eq:chargon_Lagrangian} (see \textit{SI Appendix}, Sec.~6 for details). The conserved current of this theory manifestly transforms as an SO(3) vector under pseudospin rotations, in striking agreement with the predictions presented above~\cite{Lee2018}.

\subsection{Non-zero Doping} \label{sec:finite_doping}

In this section, we turn to non-zero doping and discuss the different regions of the phase diagram of Fig.~\ref{fig:main_ideas}(a). In particular, we describe how both insulating phases considered above naturally give rise to superconductivity upon doping, with proximity to topological criticality playing the crucial role of relegating electron excitations to high energies.\footnote{This distinguishes our mechanism from those invoking proximity to a critical point in a \textit{metal}~\cite{Berg2019ARCMP}, where a fluctuating order parameter field is commonly proposed to induce electron pairing~\cite{Shibauchi2014ARCMP}.}
On the IQH side, we argue this occurs via the condensation of low-lying charge-$2e$ modes, while on the CSL side we show how superconductivity arises within a parton description of the low-lying fractionalized charge carriers.

Starting from the IQH (or Phase I), the persistence of the spin gap and reduction of the charge-$2e$ gap by topological criticality implies that doped charges enter as spin-singlet Cooper pair excitations, \textit{i.e.}, bound states of electrons. This can be seen from the effective theory (Eq.~\ref{eq:chargon_Lagrangian}) in the phase where $\psi$ is condensed.
Shifting variables $a \rightarrow a+A$ yields the coupling $2 \frac{A_0}{2\pi} \nabla \times \bm{a}$, where $\nabla\times \bm{a}=\partial_xa_y-\partial_ya_x$, which implies that charges enter as vortices of the $\psi$ condensate. In particular, by flux quantization $\int \nabla\times \bm{a} \in 2\pi\mathbb{Z}$, they are forced to enter as charge-$2e$ objects, or Cooper pairs.

Per unit cell of the triangular lattice, these pairs experience $2\pi$ external flux, twice that of their constituent electrons. Thus, magnetic translations do \textit{not} enforce degenerate minima in the energy landscape of Cooper pair excitations, unlike at generic flux fractions where the Cooper pair effective mass would also be suppressed by a narrower bosonic Hofstadter bandwidth.\footnote{The condensation of interacting Hofstadter bosons has been established even at smaller flux fractions~\cite{Oktel2007,Powell2011,Kennedy2015NatPhys}.} In fact, in Sec.~\ref{sec:numerical_evidence} we provide numerical evidence that there is a dispersive, non-degenerate $2e$ bound state. At $U/t=6$ on the $4\times 4$ torus, in subsection~\ref{sec:ED} we estimate its effective mass to be $m_{\rm Cooper} \approx 1.3\,\hbar^2/t a^2$. While this provides an order-of-magnitude estimate of the Cooper pair effective mass deep in the IQH phase, the parton theory dictates that the effective mass vanishes on approaching the critical point, with the dispersion approaching $\omega \propto |k|$. Moreover, we expect the low-lying Cooper pairs to have a characteristic size $\ell_c$ controlled by the spin gap.

We now consider doping the IQH insulator. At sufficiently low dopant density compared to $\ell_c^{-2}$ and $\xi^{-2}$, where $\xi$ is the charge-$2e$ correlation length (see \textit{SI Appendix}, Sec.~4),
we have a dilute gas of Cooper pairs with short-range repulsive interactions and small effective mass. These are expected to enter a superfluid phase, where the superfluid density increases continuously from zero as $\mu$ is tuned across the IQH-superconductor transition, as in the Bose-Hubbard model away from integer filling~\cite{Fisher1988_dilutebosegas,Fisher1989_bosehubbard}. At higher densities, it would be valuable to explore in detail whether longer-range bosonic interactions mediated by critical fluctuations can lead to crystallization~\cite{alder_superfluid_1989} or phase separation~\cite{PinesNozieres_superfluid}.

The charge and spin response of the spin-singlet IQH insulator are $\sigma_{xy}=2\cdot e^2/h$ and $\sigma^s_{xy}=2\cdot \hbar/8\pi$, respectively, with edge chiral central charge $c_{-} =2$. On condensing spin-singlet Cooper pairs, the spin quantum Hall response is unaffected since the spin sector remains gapped. However, including their superfluid response removes the quantization of charge Hall conductance, while the chiral central charge is unchanged~\cite{Kitaev2006}. This predicts edge states usually associated with the weak-pairing ``$d+id$'' superconductors~\cite{SenthilMarstonFisher1999,Moroz2017}. However, as later shown in Sec.~\ref{sec:numerical_evidence}, we are far from the weak-pairing limit, so that the symmetry of the Cooper pair wavefunction is unrelated to the topological properties and edge states of the superconductor~\cite{ReadGreen2000}.
Moreover, as we later discuss in Sec.~\ref{subsec:pairing_sym}, the terminology regarding pairing symmetry must be adapted to the present scenario of magnetic space group symmetries~\cite{shaffer_theory_2021,shaffer_unconventional_2022}.

Starting from the CSL (or Phase II), we now have fractionalized elementary excitations and must appeal to the parton description. Our analysis elucidates a remarkable connection to the mechanism of anyon superconductivity, and also the possible microscopic realization of higher-charge superconductivity in related models. We consider doping a finite charge density of $0< y\ll 1$ \textit{holes} per site, which requires that the rotor density be $\langle L_i \rangle = -y$. The constraint Eq.~\eqref{eq:U(1)_constraint} correspondingly demands a depletion of spinons relative to integer filling: $\langle n^f_{i\uparrow}+n^f_{i\downarrow}\rangle = 1-y$, which seemingly poses a problem since we expect the spin gap, and hence the spinon gap, to remain open upon light doping. Fortunately, since the spinons occupy bands with Chern number $C_\uparrow=C_\downarrow=1$, their density can be tuned by introducing additional gauge flux $n_\phi \equiv (\nabla\times\bm{a})/2\pi = -y/2$.

To evaluate the resulting physical behavior and elucidate the possibility of superconductivity, let us concentrate on hole doping and replace our rotor with a hardcore bosonic ``holon'' $b$ on each site, with $n_b =b^\dagger b \in \{ 0, 1 \}$, and write 
$c^\dagger_\sigma = b f^\dagger_\sigma$.
The Hilbert space on each site is restricted to empty and single-electron states, which both satisfy the \textit{slave-boson} constraint $n_b = 1-n_f$~\cite{song2021doping}.
Following our previous analysis but with the canonical boson $b$ rather than the rotor, integrating out $f$ leaves:
\begin{equation}
\begin{aligned}
    {\mathcal L}_{\rm doped} &= b^* \left ( i\partial_t + a_0 -A_0 \right ) b -|\vec{D}_{a - A}b|^2 - V(b) \\
    &\quad\  +\frac{2}{4\pi} a\wedge da + \frac{2}{4\pi}A_s\wedge dA_s.
\end{aligned} \label{Slave-boson-semions}
\end{equation}
This theory incorporates the feature that the density of holons is constrained to be $n_b = - 2 n_\phi$, seen from the equation of motion for $a_0$. While there are a variety of possible phases of bosons at the filling $\nu_{\rm holon} = -2$, a very natural one is the bosonic integer quantum Hall (bIQH) state with {\em even} integer Hall conductivity~\cite{LuVishwanath2012,LevinSenthil2013}, a symmetry protected topological (SPT) phase~\cite{ChenGuLiuWen} (where the protecting U(1) symmetry is taken to be the gauge ``symmetry''). The bIQH phase hosts symmetry-protected counter-propagating edge states and has no intrinsic topological order. Thus it is entirely captured by its effective response~\cite{LuVishwanath2012}, probed here by the gauge field combination $a-A$, namely $\Delta {\mathcal L}_{\rm holon} = -\frac{2}{4\pi} (a-A) \wedge d(a-A)$.
Adding this to the effective theory for Phase II we obtain the finite-doping theory in which, importantly, the dynamical Chern-Simons term cancels out:
\begin{equation}
{\mathcal L}_{\rm doped} = \frac{2}{2\pi} A\wedge da -\frac{2}{4\pi} A\wedge dA + \frac{2}{4\pi}A_s\wedge dA_s.
\label{Eq:doped}
\end{equation}
This phase therefore lacks intrinsic topological order. It has a single bulk gapless mode due to the Maxwell dynamics of $a$, representing the Goldstone mode of the spontaneously broken $\mathrm{U(1)}_c$ symmetry, absent $A$. From the Ioffe-Larkin rule for adding spinon and holon resistivity tensors, $\rho = \rho_{\rm spinon} + \rho_{\rm holon}$~\cite{IoffeLarkin1989,LeeNagaosa1990,IoffeKotliar1990,LeeNagaosaWen2006}, we also conclude that the fluid has vanishing resistivity~\cite{song2021doping}. This is because the partons have opposite Hall responses with respect to $a$:
\begin{align}
    \rho_{\rm spinon} = -\rho_{\rm holon} = \frac12 \begin{pmatrix} 0 & -1 \\ 1 & 0 \end{pmatrix}.
\end{align}
Furthermore, we see from the first term in Eq.~\eqref{Eq:doped} that a fundamental flux $\int \nabla \times \bm{a}=2\pi$ couples minimally to $A$ and binds $2e$ charge. This is therefore a superconductor of Cooper pairs, and the monopole operator that inserts unit flux of $a$ is their creation operator. In fact, Eq.~\eqref{Eq:doped} (with Maxwell terms for $a$ and $A$) can be mapped to a Ginzburg–Landau Lagrangian for a 2D superconductor~\cite{BanksLykken1990,Lykken_review}.

The Chern-Simons term for $A_s$ in Eq.~\eqref{Eq:doped} indicates a spin quantum Hall response of $\sigma_{xy}^s = 2\cdot \hbar/8\pi$. Moreover, the edge chiral central charge is $c_- = 2$ (see \textit{SI Appendix}, Sec.~3). We conclude that this superconductor has precisely the same topological and response properties as the one originating from the IQH. This motivates the minimal phase diagram shown in Fig.~\ref{fig:main_ideas}(a), where a single superconducting phase emanates from the IQH and CSL phases upon doping.

\smallskip
{\em Relation to Anyon Superconductivity---}The discussion above has a close correspondence to anyon (or more precisely {\em semion}) superconductivity, proposed in Refs.~\cite{laughlin_relationship_1988,laughlin_superconducting_1988,fetter_random_phase_1989,ChenWilczekWittenHalperin1989,WenWilczekZee1989,WenZee1989,WenZee1990,lee_anyon_1989,HosotaniChakravarty}.
These works argued that a gas of charge-$e$ semions, obtained for instance by doping a CSL, can realize a superconducting state. This problem can be mapped to that of bosons $b'$ attached to $\pi$ statistical flux, which transmutes their statistics to those of the original semions~\cite{LeeFisher1991}. In this description, the boson and flux densities are therefore related by $n_{b'} = -2 n_\phi$, exactly the relation satisfied by the slave bosons in Eq.~\eqref{Slave-boson-semions}. The analysis following that equation provides a direct link between our long-wavelength theory of the doped CSL, derived for the Hubbard-Hofstadter model, and the effective theory of semion superconductivity, most transparently its bosonic formulation developed in Ref.~\cite{LeeFisher1991}. Our analysis provides an alternative characterization of the semion superconductor, namely as a bIQH phase of the $b'$ bosons~\cite{song2021doping}.

We emphasize that a key advantage of our setup is that proximity to topological criticality in the CSL phase renders doped semions (carrying charge $e$ and no spin) energetically favorable compared to electrons. It also provides renewed motivation for studying the CSL-superconductor transition, which, to our knowledge, has not been characterized in detail in the anyon superconductivity literature. We hope that the bosonic SPT perspective presented above will offer new tools and insights for advancing its understanding~\cite{GroverVishwanath2013,LuLee2014}.

The bIQH characterization presented above also offers insight into the possible microscopic realization of charge-$Ne$ superconductivity with $N \in 2\mathbb{Z}$. Consider electrons in the fundamental representation of a flavor $\mathrm{SU}(N)$ symmetry and apply a background flux of $2\pi/N$ per unit cell. At a density of one electron per site, the non-interacting phase is an IQH insulator with $\sigma_{xy}=Ne^2/h$ where the lowest Hofstadter band is filled by all $N$ flavors. On increasing $U$, we expect to realize a CSL with $\mathrm{SU}(N)_1$ (or equivalently $\mathrm{U}(1)_N$) topological order and chiral central charge $c_{-} = N-1$~\cite{ChenHazzardReyHermele2016}. A possible critical theory is then the $\mathrm{SU}(N)$ analogue of Eq.~\eqref{eq:chargon_Lagrangian} above, represented as a complex scalar field $\chi$ coupled to a U(1) gauge field with a level-$N$ Chern-Simons term:
\begin{align} \label{eq:N_SC_Lagrangian}
    \mcL = |{\mathcal D}_{A-a}\chi|^2 - V(\chi) + \frac{N}{4\pi} a\wedge d\, a.
\end{align}
The IQH and CSL again correspond to the condensed and gapped phases for $\chi$, respectively. Upon doping, superconductivity can arise if the excess bosons enter a bIQH phase, yielding a Hall contribution $\frac{M}{4\pi}a\wedge da$ where $M$ is necessarily even~\cite{LuVishwanath2012}. Therefore, the Chern-Simons term in Eq.~\eqref{eq:N_SC_Lagrangian} may be canceled out, and a conventional superconductor can arise~\cite{BanksLykken1990}, for {\em even} integer $N$. Furthermore, these correspond to a condensate of charge-$Ne$ electron composites which are singlets under the SU($N$) and clearly can only condense for {\em even} $N$. Note that the charge-$e$ anyons in this scenario have statistical angle $\theta = \pi/N$, distinct from $\pi(1-1/N)$ in the classic anyon superconductivity scenario~\cite{ChenWilczekWittenHalperin1989}. They only agree for the special case of $N=2$, \textit{i.e.}, semion superconductivity,  considered in this work.

At $N=2$, it is tempting to view anyon superconductivity as arising from the \textit{binding} of pairs of charge-$e$ semions into bosons, that then condense. We remark however that this route would leave the remaining semionic excitations deconfined, resulting in a very different fractionalized superconductor ``SC$^*$''~\cite{SenthilFisher2000} 
that retains semionic excitations. Due to the absence of a Chern-Simons term for $a$ in Eq.~\eqref{Eq:doped}, the anyon superconductor mechanism evidently yields a superconductor of the more conventional variety~\cite{ChenWilczekWittenHalperin1989,LeeFisher1991}.

\begin{figure}
\centering
\includegraphics[width=\linewidth]{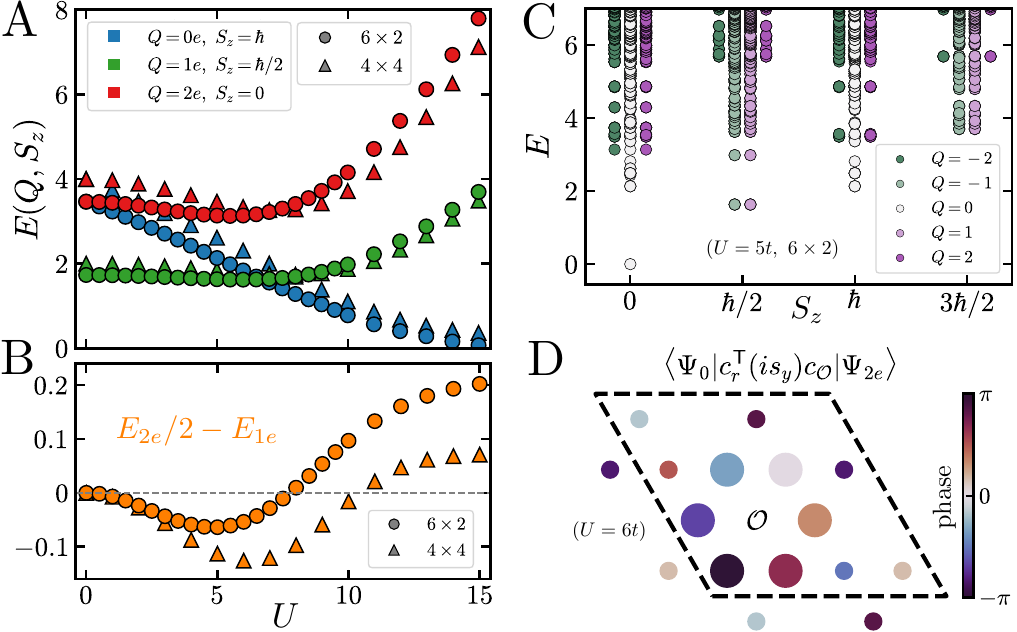}
\caption{
    (a) Exact diagonalization (ED) energies of the lowest-lying states in three charge sectors (indicated in legend) for both the $4\times 4$ site (triangular markers) and $6\times 2$ site systems (circles).
    (b) Plot of $E_{2e}/2 - E_{1e}$ vs. $U$, whose negative value indicates electron pairing.
    (c) ED energy spectrum on the $6\times 2$ site torus at $U/t=5$. The energies are arranged by total spin $S_z$ and electric charge $Q$ (indicated in legend); they organize into multiplets of the spin and pseudospin SU(2) symmetries.
    (d) Overlap of the spin-singlet pairing operator $c^\mathsf{T}_{\bm{r}}(is_y) c_{\mathcal{O}}$ between the half-filled ground state and two-electron excitation for all sites $\bm{r}$ on a $4\times 4$ site torus at $U/t=6$, with origin indicated by $\mcO$. Its magnitude and phase are specified by marker area and color, respectively.
    }
\label{fig:ED}
\end{figure}

\section{Numerical evidence for Cooper pairing} \label{sec:numerical_evidence}

In the preceding section, we presented field-theoretic arguments for the nature of the putative critical theory and its excitations, as well as the intriguing possibility of topological superconductivity upon doping near this critical point. Here, we provide explicit numerical evidence for an important precursor to superconductivity, electron pairing above the half-filled ground states.

\subsection{Exact Diagonalization} \label{sec:ED}

We first investigate the system using exact diagonalization (ED). We focus on $6\times 2$ and $4\times 4$ site systems, the latter being the largest we can access. We diagonalize the Hamiltonian at the particle-hole-symmetric point $\mu = 0$, and organize the spectrum by the charge $Q$ and spin $S_z$ quantum numbers, taking the convention where the half-filled ground state has $Q=S_z=0$ and zero energy. To construct the $6\times 2$ torus, we identify points separated by the vectors $6\bma_1$ and by $2\bma_2 - 4\bma_1$. On the $4\times 4$ system, we identify points related by $4\bma_1$ and $4\bma_2$ (see Fig.~\ref{fig:main_ideas}(b) for the definition of the lattice vectors $\bma_{1,2}$).

In Fig.~\ref{fig:ED}(a), we plot the energies of the lowest-lying excited states in the charge sectors $(Q,S_z)$ specified by $\{(0e, \hbar), (1e, \hbar/2), (2e, 0)\}$, for both the $6\times 2$ and $4\times 4$ systems. While we focus on the electron-doped side, adding holes is energetically equivalent under particle-hole symmetry. The lowest $0e$ excitation energy decreases monotonically as a function of $U$. 
On the other hand, both $1e$ and $2e$ exhibit minima at intermediate coupling: for the $6\times 2$ system, these occur at $U^\mathrm{min}_{1e}/t \simeq 6.0$ and $U^\mathrm{min}_{2e}/t \simeq 5.5$, whereas on the $4\times 4$ torus these values increase to $U^\mathrm{min}_{1e}/t \simeq 8$ and $U^\mathrm{min}_{2e}/t \simeq 7$.

The ED calculation convincingly demonstrates that the lowest-lying charge excitations are paired over a remarkably broad range of interaction strengths. Specifically, the energy $E_{2e}$ of the lowest-lying charge-$2e$ state is less than that of two individual electronic excitations, \textit{i.e.}, $E_{2e} - 2E_{1e} < 0$. We plot this quantity in Fig.~\ref{fig:ED}(b), which we find to be negative in the range $0 < U/t \leq 7.5$ for the $6\times 2$ system and over an even broader range $0 < U/t \leq 10$ for the larger $4\times 4$ system. The maximum magnitude of the pairing energy increases from $0.06t$ to $0.13t$ between the two systems, constituting $4\%$ and $7\%$ of the corresponding $1e$ energies, respectively. The location of this maximum is $U_\mathrm{max}/t \simeq 5$ for the $6\times 2$ system, increasing to $U_\mathrm{max}/t \simeq 6$ on the larger $4\times 4$ torus. Though these features occur at smaller interaction strength than the expected IQH-CSL transition point $U/t \approx 11.5$ reported by previous cylinder iDMRG studies~\cite{Kuhlenkamp2024,Divic2024}, we anticipate that $U_\mathrm{max}$ will continue to increase with increasing system size.

The numerical observation of pairing is important confirming evidence for topological criticality and the associated softening of charge-$2e$ modes. Moreover, as argued in Sec.~\ref{sec:finite_doping}, the existence of electron pairing in the IQH phase is a direct precursor to superconductivity and is likewise expected in the CSL near criticality. To estimate the Cooper pair effective mass, crucial to phase stiffness, we thread small flux $\varphi$ through the torus and compute the change in the charge-$2e$ energy $E_{2e}$.
For concreteness, we fix $U/t=6$ on the $4\times 4$ torus, where pairing is maximal, though on larger systems we expect this point to reside deep in the IQH phase (see Refs.~\cite{Kuhlenkamp2024,Divic2024}).
At small $\varphi$, we obtain a quadratic fit $E_{2e}(\varphi)/t \approx 5.01 (\varphi/2\pi)^2$. Threading $2\pi$ flux shifts the total \textit{two}-electron momentum $\bm{q}$ by $2/L$ times a primitive reciprocal vector of magnitude $4\pi/a\sqrt{3}$, where $L=4$ is the length of the torus. Matching the dispersion to $\hbar^2 q^2/2m_\mathrm{Cooper}$ with $\varphi/2\pi = q a\sqrt{3}/2\pi$, we estimate $m_\mathrm{Cooper} \approx 1.3\, \hbar^2/ta^2$ at the chosen system size and interaction strength, providing an order-of-magnitude estimate for the pair effective mass in the IQH phase.

We note that the pairing energy $E_{2e}/2 - E_{1e}$ 
cannot be positive in the thermodynamic limit because it is always possible to create a well-separated pair of $1e$ excitations with energy equal to that of independent electrons.
Its taking positive values for $U/t > 7.5$ and $U/t > 10$ on the $6\times 2$ and $4\times 4$ systems, respectively, is therefore a finite size effect. Given that the interactions are repulsive, the observed (negative) electron pairing energy at smaller $U$ has no similarly-compelling finite-size explanation, thus pointing to topological criticality as its origin.
Furthermore, the fact that pairing extends down to seemingly arbitrarily-small values (probed down to $U/t=10^{-3}$) suggests a weak-coupling origin complementary to pairing induced strictly by proximity to criticality, which we further elucidate in Sec.~\ref{eq:pairing_valence_edge} by a small-$U$ perturbative calculation on much larger systems.

Each of the systems under consideration possesses the $\mathrm{SO(4)}$ symmetry explicated in Sec.~\ref{sec:model_symmetries}. Accordingly, we observe that the ED spectrum organizes into multiplets of the spin and pseudospin $\mathrm{SU(2)}$ symmetries. This is displayed in Fig.~\ref{fig:ED}(c) for the $6\times 2$ torus at $U/t=5$, in which eigenstates with fixed spin $S_z$ but different total charge $Q$ populate multiplets of the pseudospin $\mathrm{SU(2)}_c$.
For both system sizes, the $(Q,S_z)=(0e,\hbar)$ state in Fig.~\ref{fig:ED}(a) belongs to a pseudospin-singlet spin-triplet irreducible representation, while the $(Q,S_z)=(2e,0)$ branch is a pseudospin-triplet spin-singlet.

We remark that pairing was not observed on the smaller $4\times 2$ torus, nor on the $6\times 2$ system with the identification $\bm{r} \sim \bm{r} + 2\bm{a}_2 \sim \bm{r} + 6\bm{a}_1$, though we recall that pairing was observed at $6\times 2$ under the boundary identification specified earlier. Nonetheless, it is suggestive that pairing is present on the largest accessible system, the $4\times 4$ torus studied above. Verifying pairing at larger system sizes, employing approximate methods beyond ED, is an important task for future work. In the next section, we take one step forward by establishing electron pairing in a cylindrical geometry of finite width but \textit{infinite} length using tensor network methods.

\begin{figure*}
    \centering
    \includegraphics[width = 17.8cm]{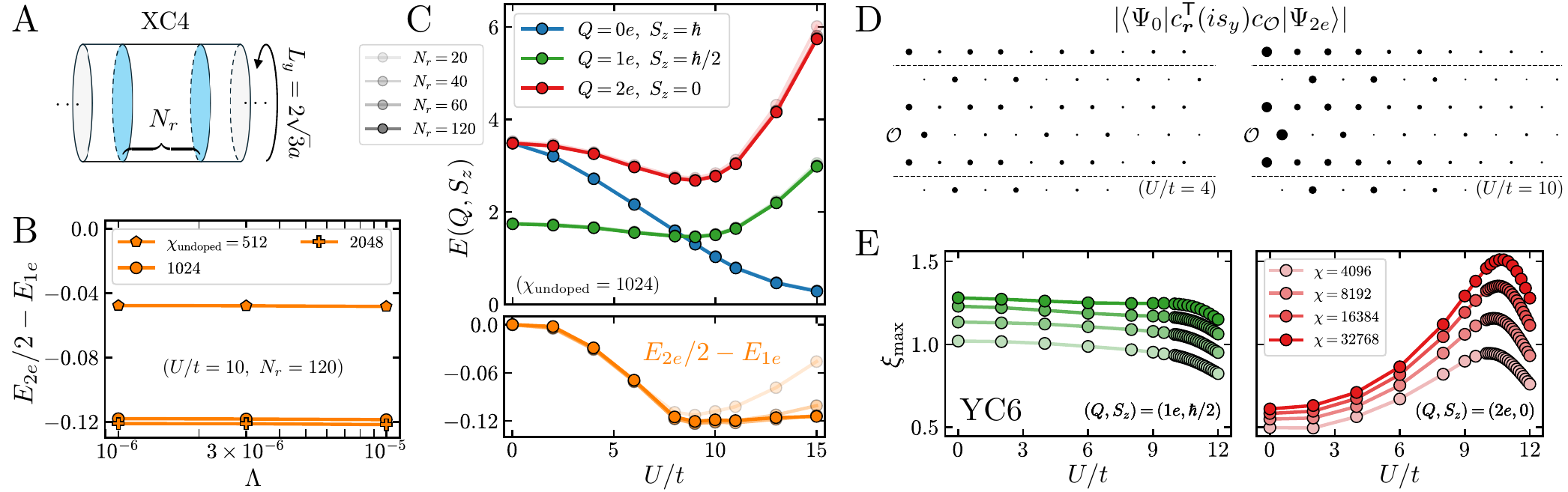}
    \caption{
    (a) Illustration of the infinite-length $\mathrm{XC}4$ cylinder with circumference $L_y=2\sqrt{3}a$, where $a$ is the lattice spacing. The ``segment'' DMRG method consists of optimizing a finite number of tensors (describing $N_r$ cylinder rings) sandwiched between two semi-infinite MPS environments derived from a reference (half-filled) ground state with bond dimension $\chi_\text{undoped}$.
    (b) Electron pairing energy $E_{2e}/2-E_{1e}$ vs. maximum magnitude $\Lambda$ of discarded singular values in the finite segment, with legend indicating $\chi_\text{undoped}$.
    (c) Upper panel: energies of excitations with charges $(Q,S_z)$ (indicated in legend) vs. Hubbard $U$, with fixed $\chi_\text{undoped}=1024$ and $\Lambda = 10^{-5}$. 
    The legend indicates $N_r$ (increasing light to dark).
    Lower panel: electron pairing energy vs. $U$.
    (d) Magnitude of the pair wavefunction at both $U/t=4$ and $U/t=10$, computed from respective iDMRG ground states with $\chi_\text{undoped}=1024$ and segment excitations with $N_r=120$. The two dotted lines are related under the cylinder identification; outside points are periodic images.
    (e) Correlation lengths of the half-filled YC6 ground state in the $1e$ (left panel) and $2e$ (right) sectors, in units of cylinder rings, as a function of bond dimension (increasing light to dark) and $U$.
    }
    \label{fig:segmentDMRG}
\end{figure*}

\subsection{Segment DMRG}

Here we employ the ``infinite boundary condition'' technique~\cite{Phien2012,ZaletelMongPollmann2013, Phien2013,Zauner2015}, referred to henceforth as the ``segment'' DMRG method~\cite{ZaletelMongPollmann2013,Chatterjee2022,tenpy}, to obtain excited states and their energies. Starting from an infinite MPS approximation to the ground state at half-filling, we allow the tensors of the excited state MPS to differ on a segment spanning $N_r$ cylinder rings. DMRG is then used within this variational class to minimize the excitation energy for fixed quantum numbers $Q, S_z$, and circumferential momentum $k_y$ relative to the ground state. The segment DMRG method lets us probe finite-charge excitations above the half-filled ground state without introducing a physical edge.

For the XC4 cylinder considered here and shown schematically in Fig.~\ref{fig:segmentDMRG}(a), we identify points separated by $4\bma_2 - 2\bma_1 \propto \hat{y}$ around the circumferential direction (with circumference $2\sqrt{3}a$), but take the cylinder to have infinite length, with $\bma_1$ parallel to the cylinder axis~\cite{Szasz2020}. Each cylinder ring consists of two spinful lattice sites.
We employ a gauge in which the magnetic unit cell consists of two sites, with magnetic Bravais vectors given by $\bma_1$ and $2\bma_2 - \bma_1$ (see \textit{SI Appendix}, Sec.~1C). This results in two distinct momenta $k_y$, which label the eigenvalues of the circumferential translation $(T_2)^2 T_1^\dagger$. To obtain the parent half-filled states, we use the iDMRG algorithm~\cite{White1992,White1993,mcculloch2008}, and conserve the charges $(Q,S_z,k_y)$ for both the ground state and the segment excitation calculations.

Accurate results require converging in three separate parameters: the bond dimension $\chi_\mathrm{undoped}$ of the half-filled ground state, the bond dimension of tensors within the variational segment---which we parameterize by the maximum magnitude $\Lambda$ of the discarded singular values---and the number of cylinder rings $N_r$ in the segment. In Fig.~\ref{fig:segmentDMRG}(b), we plot the pairing energy (for $U/t=10$ and $N_r=120$ rings) as a function of $\Lambda$ and parent state bond dimension. The energy is found to depend more strongly on the latter, but shows little quantitative difference between $\chi_\mathrm{undoped}=1024$ and $2048$, permitting us to choose the less expensive option at other interactions $U$.
We focus on the XC4 cylinder because it is the largest system on which we can obtain $\lesssim 0.01t$ error in the energy using available resources, and because the smaller YC3 cylinder does not exhibit pairing, presumably due to insufficient cylinder width.

In Fig.~\ref{fig:segmentDMRG}(c), we plot the excitation energies in the charge sectors $(Q,S_z)$ given by $\{(0e, \hbar), (1e, \hbar/2), (2e, 0)\}$ as a function of Hubbard interaction $U$ and the segment length $N_r$ (increasing with shading, light to dark), fixing $\chi_\text{undoped}=1024$. For each $(Q,S_z)$, we initialize in each momentum sector $k_y$ and take the minimum energy. We remark that the energy curves bear strong resemblance to those obtained from ED in Fig.~\ref{fig:ED}(a). For the DMRG, we observe that the smaller-$U$ simulations exhibit more rapid energetic convergence in $N_r$. Both the $1e$ and $2e$ energies monotonically decrease from $U/t=0$ to a minimum at $U/t\approx 9$ and increase rapidly thereafter. Consistent with ED, their energies indicate electron pairing $E_{2e}/2 - E_{1e} < 0$ over a broad range of interaction strengths, from $U=0$ to at least $U/t=15$. The maximum pairing strength of $0.12t$ occurs at $U/t=9$, near the putative critical point discussed in Sec.~\ref{sec:criticality_half_filling}, constituting $8.3\%$ of the corresponding $1e$ energy, though the data in fact exhibits a plateau in the pairing energy beginning at this $U$.

The behavior of the charge-neutral $S_z = \hbar$ excitation branch (see Fig.~\ref{fig:segmentDMRG}c) strongly resembles the corresponding energy curve in ED, where we identified it as belonging to a spin-triplet, pseudospin-singlet representation. In the segment DMRG data, this excitation energy decreases gradually from its expected value of $2\sqrt{3}t$ at $U=0$ to zero at $U/t > 15$. This agrees with the results from ED for both the $6\times 2$ and $4\times 4$-site systems, shown in Fig.~\ref{fig:ED}(a), where this gap similarly closes at $U/t > 15$. Agreement between these three geometries suggests this feature persists in the thermodynamic limit, where it may be associated with a transition to the $120^\circ$ antiferromagnetic phase, setting an upper-bound on the extent of the putative CSL phase.

Though our explicit DMRG pairing calculations are limited to the XC$4$ cylinder, indirect evidence for the persistence and possible enhancement of pairing on wider cylinders is provided by the behavior of the transfer matrix spectrum associated with the translation-invariant half-filled parent state~\cite{Kuhlenkamp2024,Divic2024}. The dominant transfer matrix eigenvalues, resolved by charge sector $(Q,S_z,k_y)$, relate to ground state connected correlation lengths of operators carrying those charges~\cite{Zauner2015NJPh,SCHOLLWOCK201196}. In the $\text{YC-}L_y$ cylinder sequence~\cite{Szasz2020}, the correlation length associated with spin-singlet $2e$ excitations becomes more pronounced as $L_y$ increases from $L_y=3$ to $6$ (see \textit{SI Appendix}, Sec.~4). In Fig.~\ref{fig:segmentDMRG}(e), we showcase the $2e$ and $1e$ correlation lengths on the YC6 cylinder, where the pronounced peak of $\xi_{2e}$ and its prominence over $\xi_{1e}$ suggests the persistence of electron pairing on this wider cylinder.
The same pattern holds when comparing the $\text{XC}4$ and $\text{XC}6$ cylinders. In all cases, by threading flux if necessary, we ensure that the fluxes penetrating the cylinder rings are consistent with the particle-hole SO(4) symmetry of Sec.~\ref{sec:model_symmetries}.

\subsection{Pairing Symmetry} 
\label{subsec:pairing_sym}

In this section, we discuss the symmetry of the charge-$2e$ paired states obtained in ED and DMRG. We argue that these low-energy excitations are spin-singlet and energetically non-degenerate, carrying odd angular momentum under site-centered rotations and even angular momentum under bond-centered rotations.

On the $4\times 4$ system studied in ED, the uniqueness of the lowest-energy charge-$2e$ excited state for all $U$ implies that it is spin-singlet and transforms in a one-dimensional irreducible representation of the magnetic space group generated by $T_1, C_6$. We find the pair carries momentum $T_j|\Psi_{2e}\rangle=-|\Psi_{2e}\rangle$ for all nearest-neighbour translations $T_j$. 
Counterintuitively, this is in fact the unique momentum which is rotation-symmetric: invariance under $C_6$ requires each $T_j$ to have the same eigenvalue, while Eq.~\eqref{eq:mag_algebra_relations} requires $T_5 T_3 T_1 = -1$ when acting on a pair, which together imply $T_j = -1$.
Moreover, we find that $C_{2}|\Psi_{2e}\rangle = -|\Psi_{2e}\rangle$, so that the state has odd angular momentum with respect to site-centered rotations.
However, for each \emph{bond}-centered rotation $C^\mathrm{bond}_{2} = T_j {C}_{2}$, we therefore have $C^\mathrm{bond}_{2}|\Psi_{2e}\rangle = + |\Psi_{2e}\rangle$, which has invariant meaning since $(C^\mathrm{bond}_{2})^2 = 1$ in every fermion number sector. Thus, the pair has even angular momentum with respect to all bond-centered rotations.\footnote{All other bond-centered two-fold rotations are related by the square of some translation operator and thus have the same eigenvalue.}

By anti-symmetry of the wavefunction, spin-singlet pairing should therefore be allowed between sites related by $C^\mathrm{bond}_{2}$ but disallowed when they are related by a site-centered rotation $C^\mathrm{site}_{2}$.
This is exactly borne out in the $4\times 4$ ED numerical data. In Fig.~\ref{fig:ED}(d), we plot the spin-singlet ``pair wavefunction''
$\langle \Psi_0|c^\mathsf{T}_{\bmr}(is_y)c_{\mcO}|\Psi_{2e}\rangle$ at $U/t=6$, where $\Psi_0$ is the ground state, $\mcO$ is the origin, and $c^\mathsf{T}$ denotes the transpose of the column vector of electron annihilation operators.
We observe that the wavefunction vanishes when $\bm{r}$ is related to $\mcO$ by some $C^\mathrm{site}_{2}$, in agreement with the above arguments. Moreover, the pair is well-localized, with nearest-neighbour pairing having much larger magnitude than next-nearest-neighbour pairing.
Though the phase winding in Fig.~\ref{fig:ED}(d) suggests the pairing symmetry is ``$p+ip$'', we caution that the $C_3$ eigenvalue of the charge-$2e$ excited state does not have obvious invariant meaning as it can be modified by the redefinition $C_3 \to (e^{2\pi i/3})^{N_F} C_3$. Nonetheless, on the $4\times 4$ torus, we have unambiguously confirmed spin-singlet pairing, carrying odd (even) angular momentum under site-centered (bond-centered) rotations.

On the XC4 cylinder, three-fold rotation symmetry is explicitly broken. Nonetheless, the segment DMRG pair excitations exhibit the same pairing symmetry as above.
We verify this explicitly by computing the pair wavefunction, choosing $\mcO$ at the center of the segment.\footnote{The inversion center $\mcO$ is approximate since the segment contains a finite, even number of rings.} We find it is odd under $\bm{r}\to C_2\bm{r}$ and that the spin-triplet overlap is several orders of magnitude smaller than spin-singlet. Moreover, as shown in Fig.~\ref{fig:segmentDMRG}(d) for $U/t=4$ and $10$, we find that the pair wavefunction vanishes approximately (error is induced by finite $N_r$) when $\bm{r}$ and $\mcO$ are related by some ${C}^\text{site}_2$, which indicates ${C}^\text{site}_2|\Psi_{2e}\rangle = -|\Psi_{2e}\rangle$.
Similarly, we conclude that ${C}^\text{bond}_2|\Psi_{2e}\rangle = +|\Psi_{2e}\rangle$.
These plots also show that the size of the excitation decreases with $U$. Finally, since we conserve momentum around the cylinder, our excited states are labeled by their eigenvalue under the circumferential translation ${T} = (T_2)^2 T_1^\dagger$. For all $U$, the paired eigenstate satisfies ${T}|\Psi_{2e}\rangle = -|\Psi_{2e}\rangle$,
consistent with our expectation $T_1 = T_2 = -1$.

We anticipate that our pairing symmetry results will help guide future numerical investigations of pairing and superconductivity in this model, especially those employing techniques that operate within a fixed charge sector, such as the ED and DMRG performed in this work.

\subsection{Pairing at the conduction band edge} \label{eq:pairing_valence_edge}

Both our ED (Fig.~\ref{fig:ED}) and DMRG (Fig.~\ref{fig:segmentDMRG}) numerics indicate that pairing extends down to small values of $U$. Although our analytical discussion has focused largely on the intermediate-coupling regime near topological criticality, the regime of perturbatively-weak interactions allows for an independent check of the existence and symmetry of bound pairs in our model, which we fix here to be spin-singlet. 
To second order in many-body perturbation theory, renormalization of the two-body scattering vertex is described by the following diagrams~\cite{KohnLuttinger1965,MaitiChubukov2013,Kagan2014JETP}:
\begin{equation} \label{eq:FiveKLDiagrams}
\begin{tikzpicture}[baseline=0cm,scale=0.8]
\draw (-0.75,0.5) -- (0.75,0.5); \draw (-0.75,-0.5) -- (0.75,-0.5); 
\draw[snake it] (-0.33,0.5) -- (-0.33,-0.5); \draw[snake it] (0.33,0.5) -- (0.33,-0.5);
\end{tikzpicture} , \, 
\begin{tikzpicture}[baseline=0cm,scale=0.8]
\draw (-0.75,0.5) -- (0.75,0.5); \draw (-0.75,-0.5) -- (0.75,-0.5); 
\draw[snake it] (-0.33,0.5) -- (0.33,-0.5); \draw[snake it] (0.33,0.5) -- (-0.33,-0.5);
\end{tikzpicture} , \, 
\begin{tikzpicture}[baseline=0cm,scale=0.8]
\draw (-0.75,-0.5) -- (0.75,-0.5); \draw (-0.75,0.5) -- (-0.33,0.5) -- (0,0) -- (0.33,0.5) -- (0.75,0.5); 
\draw[snake it] (0,-0.5) -- (0,0); \draw[snake it] (0.33,0.5) -- (-0.33,0.5);
\end{tikzpicture} , \, 
\begin{tikzpicture}[baseline=0cm,scale=0.8]
\draw (-0.75,0.5) -- (0.75,0.5); \draw (-0.75,-0.5) -- (-0.33,-0.5) -- (0,0) -- (0.33,-0.5) -- (0.75,-0.5); 
\draw[snake it] (0,0.5) -- (0,0); \draw[snake it] (0.33,-0.5) -- (-0.33,-0.5);
\end{tikzpicture} , \, 
\begin{tikzpicture}[baseline=0cm,scale=0.8]
\draw (0,0) circle (0.25); 
\draw (-0.75,0.5) -- (0.75,0.5); \draw (-0.75,-0.5) -- (0.75,-0.5); 
\draw[snake it] (0.,0.5) -- (0.,0.25); \draw[snake it] (0.,-0.5) -- (0,-0.25);
\end{tikzpicture} 
\end{equation}
where curvy and straight lines denote interaction events and single-particle propagators, respectively. As shown in Ref.~\cite{Crepel2022} in the context of superconductivity mediated by local repulsion~\cite{crepel2021new,crepel2022unconventional}, some of these diagrams vanish given a specific bare interaction kernel. 
For our model at one electron per site and zero temperature, we show in \textit{SI Appendix}, Sec.~7 that only the ``cross'' diagram contributes to the Cooper vertex in the spin-singlet channel. 
Restricting our attention to the four degenerate single-particle states at the conduction band edge, labeled by their momenta $\kappa,\kappa'$ and degenerate band index $n\in\{1,2\}$ (see Fig.~\ref{fig:main_ideas}c), we find that this diagram only mediates pairing in the channel
\begin{equation}
\begin{aligned}
\hat{\Delta} &= c_{\kappa',\uparrow,2} c_{\kappa,\downarrow,1} - c_{\kappa',\downarrow,2} c_{\kappa,\uparrow,1}  \\ 
&\qquad - c_{\kappa,\uparrow,2} c_{\kappa',\downarrow,1} + c_{\kappa,\downarrow,2} c_{\kappa',\uparrow,1}.
\end{aligned}
\end{equation}
Consistent with ED and DMRG, this spin-singlet pair has odd angular momentum with respect to site-centered rotations.

\section{Discussion}
\label{sec:Discussion}

A central message of this work is that topological superconductivity can emerge in a regime with both strong repulsive interactions and broken time reversal symmetry. Electron pairing was investigated and was clearly observed in our numerical calculations in the present Hofstadter-Hubbard model. It is strongest in the regime of intermediate interaction strength close to the putative IQH-CSL critical point~\cite{Kuhlenkamp2024,Divic2024} and remarkably remains non-zero down to perturbatively-weak interactions. The pairs are spin-singlet, with odd (even) angular momentum under site-centered (bond-centered) rotations.

These numerical observations support the principal idea that proximity to topological quantum criticality can offer a robust route to superconductivity, even in remarkably simple settings. The crucial ingredient here is that the topological transition is associated with the closing of the charge gap,  while the spin gap remains open.
From the IQH-CSL critical theory formulated for the present Hofstadter-Hubbard model, we argued that doping naturally leads to a topological superconductor. On the CSL side of the phase diagram, this relates to the storied mechanism of semion superconductivity. However, given the broad extent around criticality where superconductivity  is anticipated, it is not strictly necessary for this mechanism that the topological critical point or the CSL be accessed in a given model. For instance,
superconductivity could emerge upon doping the IQH well away from criticality, even in a scenario where the CSL is entirely subsumed by a conventional magnetically-ordered phase, or if the transition exists but is weakly first-order.
Likewise, we do not expect $C_2$ breaking or deviation from particle-hole symmetry to play a significant role away from the weak-coupling regime.

We now list key areas for future exploration. A more complete characterization of the quantum critical point and its associated conformal field theory would further elucidate the likely-universal nature of superconducting onset upon doping the IQH, CSL, and the critical point itself. It may also shed light on potential competition with conventional~\cite{alder_superfluid_1989} and fractionalized~\cite{jiang_holon_2017,Jiang2024} crystalline phases at non-zero doping. Another immediate next step is numerically verifying the existence of the superconducting ground state in this model, for instance with cylinder DMRG~\cite{Motruk2016,JiangDevereaux_science2019,Qin2020,SahayDivic} or other suitable methods. Furthermore, to distinguish features unique to our model from those of the general theoretical scenario near criticality, it will be necessary to compare with a broader class of models that relax various symmetries. Such results should also be compared to existing numerical observations of superconductivity in related models on the triangular lattice~\cite{ZhuChen2023}, most notably chiral $t$-$J$ models with real hoppings~\cite{Jiang2020,Huang2022PRX,Huang2023PRL,Zhu2023}, where the charge fluctuations integral to topological criticality are absent at half filling.

Next, we highlight promising experimental realizations in moir\'e materials. Triangular lattices formed by TMD sheets with a moir\'e potential offer both valley and layer ``pseudospin'' degrees of freedom~\cite{wu2019topological,zhang2021electronic,crepel2023topological,zheng2023localization,zhang2023approximate,crepel2024attractive}. As proposed in Refs.~\cite{Zhang2021,Kuhlenkamp2024}, the Hofstadter-Hubbard model considered here could be realized by applying a perpendicular magnetic field to such a system with a tens-of-nanometer moir\'e lattice constant. The field would conveniently polarize the valley (\textit{i.e.}, true spin) degree of freedom due to the Zeeman effect, yielding the effective model of Sec.~\ref{sec:model_symmetries}, with layer pseudospin playing the role of spin. Further numerical studies of this model and its deformations would help guide the experimental exploration of criticality and superconductivity in these moir\'e platforms.

Finally, the recent observation of the fractional quantum anomalous Hall effect in twisted bilayer $\mathrm{MoTe}_2$~\cite{Cai2023Nature,Park2023Nature,Zeng2023,XuPRX2023} and rhombohedral multilayer graphene~\cite{Lu2024Nature}, as well as fractional Chern insulators in both magic-angle graphene at weak magnetic field~\cite{Xie2021Nature} and in the field-induced Chern bands of Bernal bilayer graphene aligned with hexagonal boron nitride~\cite{Spanton2018}, motivate exploring the behavior of charged anyons subject to substantial lattice effects, as we have in this work. Further, a recent experiment on quadrilayer rhombohedral graphene demonstrates that chiral superconductivity is possible even at strong magnetic fields~\cite{han_signatures_2025}, a hopeful sign for the class of mechanisms proposed here, motivating the study of routes to pairing and superconductivity in the strong time-reversal-breaking regime.

\bigskip
\begin{acknowledgments}
We acknowledge helpful discussions with S. Anand, S. Chatterjee, R. Fan, D. Guerci, Y-C. He, C. Kuhlenkamp, E. Lake, D-H. Lee, Z.D. Shi, E.M. Stoudenmire, and Z. Weinstein. We are especially grateful to Y-H. Zhang, R. Verresen, and E. Altman for insightful discussions.
S.D. and M.Z. were supported by the U.S. Department of Energy, Office of Science, National Quantum Information Science Research Centers, Quantum Systems Accelerator (QSA).
A.V. was supported by the Simons Collaboration on Ultra-Quantum Matter, which is a grant from the Simons Foundation (618615, A.V.) and by NSF DMR-2220703.
A.M. was supported in part by Programmable Quantum Materials, an Energy Frontier Research Center funded by the U.S. Department of Energy (DOE), Office of Science, Basic Energy Sciences (BES), under award DE-SC0019443.
This research is funded in part by the Gordon and Betty Moore Foundation’s EPiQS Initiative, Grant GBMF8683 to T.S.
Calculations were performed using the TeNPy Library~\cite{tenpy,tenpy_v1}.
The Flatiron Institute is a division of the Simons Foundation.

\smallskip
\noindent\textit{Note added.} As this manuscript was being completed, Refs.~\cite{KimTimmelJuWen2024,ShiSenthil2024} appeared, which share some overlap with the present work.
\end{acknowledgments}

\endgroup

\bibliographystyle{apsrev4-1} 
\bibliography{main} 

\definecolor{shadecolor}{gray}{0.9}

\clearpage

\onecolumngrid
\begin{center}
\textbf{\large Supporting Information: ``Anyon Superconductivity from Topological Criticality in a Hofstadter-Hubbard Model''}
\end{center}

\setcounter{equation}{0}
\setcounter{figure}{0}
\setcounter{table}{0}
\makeatletter
\setcounter{secnumdepth}{2}
\renewcommand{\thefigure}{S\arabic{figure}}
\renewcommand{\thesection}{S\Roman{section}}
\renewcommand{\thesubsection}{S\Roman{subsection}}
\renewcommand{\bibnumfmt}[1]{[S#1]}
\setcounter{section}{0}

\newcommand{\toclesslab}[3]{\bgroup\let\addcontentsline=\nocontentsline#1{#2\label{#3}}\egroup}

\appendix

\setcounter{figure}{0} 
\renewcommand{\thefigure}{A\arabic{figure}}
\renewcommand{\theHfigure}{A\arabic{figure}}

\tableofcontents

\section{Coordinates, Gauge Choice and Symmetries} \label{app:gauge_choices}

\subsection{Triangular lattice coordinates}

Here, we specify the Bravais vectors of the underlying triangular lattice (see Fig.~1(b) of the main text):
\begin{align}
\bm{a}_{1} = \hat{x}, \qquad 
\bm{a}_{2} = \frac{\hat{x}}{2} + \frac{\hat{y}\sqrt{3}}{2}
\end{align}
where we've set the nearest-neighbour lattice bond length to unity, $a=1$. The lattice points are then given by
\begin{align} \label{eq:r_as_n1_n2}
\bm{r}=n_{1}\bm{a}_{1}+n_{2}\bm{a}_{2},\qquad n_{i}\in\mathbb{Z},
\end{align}
We take the reciprocal lattice vectors to be
\begin{align}
\bm{b}_{1} = 2\pi\hat{x} - \frac{2\pi}{\sqrt{3}} \hat{y}, \qquad 
\bm{b}_{2} = \frac{4\pi}{\sqrt{3}} \hat{y}.
\end{align}

\begin{figure*}
    \centering
    \includegraphics[width = 494 pt]{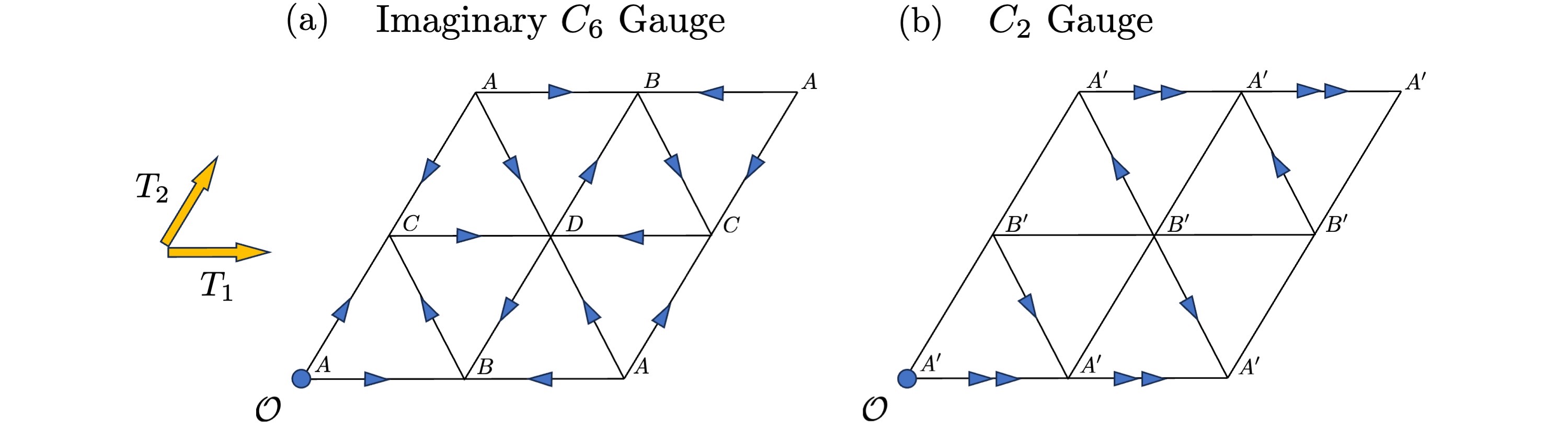}
    \caption{Two gauge choices for the hopping Hamiltonian, in both cases giving rise to $\Phi_\triangle=\pi/2$ magnetic flux per triangular plaquette. (a) Hoppings in the ``Imaginary $C_6$'' gauge, which is manifestly $C_6$ symmetric about the origin [bottom left marked by $\mcO$]. The origin is chosen to coincide with the $A$ sublattice (other sublattices $B,C,D$ are also marked). Each arrow indicates a Peirel's phase of $+i$ in the direction of the arrow. This gauge choice is identical to that displayed in Fig.~1(b) of the main text, except there the origin is in the middle of the image.
    (b) Hoppings in the ``$C_2$'' gauge. All hoppings are either imaginary or real, with a single arrow again indicating a Peirel's phase of $+i$ in the direction of the arrow. There are two magnetic sublattices, $A'$ and $B'$. For both (a) and (b): $T_1$ and $T_2$ (yellow arrows) denote translation symmetry in the unit directions $\bma_1$ and $\bma_2$, respectively. In App.~\ref{app:imag_C6_gauge} we provide the explicit form of $T_1$ and $T_2$ in the Imaginary $C_{6}$ gauge.
    }
    \label{fig:two_gauges}
\end{figure*}

\subsection{Imaginary $C_{6}$ gauge} \label{app:imag_C6_gauge}

Fig.~\ref{fig:two_gauges}(a) provides an electronic gauge in which the hoppings are purely imaginary \textit{and} in which the $C_{6}$ symmetry acts in its bare form in real space:
\begin{align} \label{eq:C6}
    C_6 c^\dag_{\bmr} C_6^{-1} = c^\dag_{C_6 \bmr},
\end{align}
provided that the (infinite 2D or torus) geometry is such that $C_6$ is well-defined. Thinking of a torus as a set of equivalence classes of points on the infinite plane, a well-defined spatial operation must map between equivalence classes. For example, by direction inspection, one can see that $C_6$ is ill-defined on the $4\times 2$ torus.

By inspecting the phases of the hoppings in Fig.~\ref{fig:two_gauges}(a), we read off
\begin{align} \label{eq:T1_T2}
    T_1 c^\dag_{\bmr} T_1^{-1} = i(-1)^{n_1 + n_2} c^\dag_{\bmr + \bma_1}, \qquad T_2 c^\dag_{\bmr} T_2^{-1} = i(-1)^{n_2} c^\dag_{\bmr + \bma_2},
\end{align}
where $\bm{r}=n_{1}\bm{a}_{1}+n_{2}\bm{a}_{2}$ is the lattice point before the transformation. To see the former, note that only the hoppings parallel to $\bma_1-\bma_2$ are left invariant by bare $T_1$, \textit{i.e.}, the hoppings in the $\bma_1$ and $\bma_2$ directions pick up a minus sign that needs to be corrected. For the latter, bare $T_2$ instead negates the hoppings in the $\bma_2$ and $\bma_1 - \bma_2$ directions. The factors $i$ are chosen so that two-fold translation is
a pure shift:
\begin{align} \label{eq:twosite_pure_shift}
(T_{j})^{2}c_{\bm{r}}^{\dagger}(T_{j})^{-2}=c_{\bm{r}+2\bm{a}_{j}}^{\dagger}.
\end{align}
But more importantly, the definitions Eq.~\eqref{eq:C6} and Eq.~\eqref{eq:T1_T2} are chosen to be consistent with all the algebraic conditions laid out in Sec.~1 of the main text.

As a non-trivial check of correctness of these translations, we can show that they correctly anti-commute:
\begin{align}
    (T_1\circ T_2)\cdot c^\dag_{\bmr} &= -T_1 T_2 c^\dag_{\bmr} T_2^{-1} T_1^{-1} \\
    &= -(-1)^{n_2} T_1 c^\dag_{\bmr + \bma_2} T_1^{-1} \\ 
    &= (-1)^{n_2} (-1)^{n_1 + n_2} c^\dag_{\bmr + \bma_2 + \bma_1}
\end{align}
compared to
\begin{align}
    (T_2 \circ T_1)\cdot c^\dag_{\bmr} &= -T_2 T_1 c^\dag_{\bmr} T_1^{-1} T_2^{-1} \\
    &= -(-1)^{n_1 + n_2} T_2 c^\dag_{\bmr + \bma_1} T_2^{-1} \\
    &= - (-1)^{n_1 + n_2} (-1)^{n_2} c^\dag_{\bmr + \bma_1 + \bma_2},
\end{align}
so that indeed $T_2 T_1 = (-1)^{N_F} T_1 T_2$.

The phases can be written in a more geometric form:
\begin{align} \label{eq:Ts_in_C6gauge}
    T_1 c^\dag_{\bmr} T_1^\dagger = i e^{i \bmr\cdot(\bmb_1 + \bmb_2)/2} c^\dag_{\bmr + \bma_1}, \qquad T_2 c^\dag_{\bmr} T_2^\dagger = i e^{i \bmr\cdot \bmb_2/2} c^\dag_{\bmr + \bma_2}.
\end{align}
From these, we can also obtain
\begin{align}
    (T_{1}^{\dagger}T_{2})c_{\bm{r}}^{\dagger}(T_{1}^{\dagger}T_{2})^{\dagger}=e^{-i\bm{r}\cdot\bm{b}_{1}/2}c_{\bm{r}-\bm{a}_{1}+\bm{a}_{2}}^{\dagger},
\end{align}
and by inverting the above three,
\begin{align}
    T_{1}^{\dagger}c_{\bm{r}}^{\dagger}T_{1} &= ie^{-i\bm{r}\cdot(\bm{b}_{1}+\bm{b}_{2})/2}c_{\bm{r}-\bm{a}_{1}}^{\dagger} \\
    T_{2}^{\dagger}c_{\bm{r}}^{\dagger}T_{2} &= ie^{-i\bm{r}\cdot\bm{b}_{2}/2}c_{\bm{r}-\bm{a}_{2}}^{\dagger} \\
    (T_{1}^{\dagger}T_{2})^{\dagger}c_{\bm{r}}^{\dagger}(T_{1}^{\dagger}T_{2}) &= -e^{i\bm{r}\cdot\bm{b}_{1}/2}c_{\bm{r}+\bm{a}_{1}-\bm{a}_{2}}^{\dagger}.
\end{align}
From this we learn that $C_{6}T_{1}C_{6}^{\dagger} = T_2$:
\begin{align}
(C_{6}T_{1}C_{6}^{\dagger})c_{\bm{r}}^{\dagger}(C_{6}T_{1}C_{6}^{\dagger})^{\dagger} & =(C_{6}T_{1})c_{C_{6}^{-1}\bm{r}}^{\dagger}(C_{6}T_{1})^{\dagger}\\
 & = iC_{6}e^{iC_{6}^{-1}\bm{r}\cdot(\bm{b}_{1}+\bm{b}_{2})/2}c_{C_{6}^{-1}\bm{r}+\bm{a}_{1}}^{\dagger}C_{6}^{\dagger}\\
 & = ie^{i\bm{r}\cdot C_{6}(\bm{b}_{1}+\bm{b}_{2})/2}c_{\bm{r}+C_{6}\bm{a}_{1}}^{\dagger}\\
 & = ie^{i\bm{r}\cdot \bm{b}_{2}/2}c_{\bm{r}+\bm{a}_{2}}^{\dagger}\\
 & =T_{2}c_{\bm{r}}^{\dagger}T_{2}^{\dagger}.
\end{align}
Similarly, it can be shown that $T_3 = C_6 T_2 C_6^\dagger = (-i)^{N_F} T_1^\dagger T_2$ and $C_6 T_3 C_6^\dagger = T_1^\dagger,$ which is altogether consistent with $C_2 T_j C_2 = (T_j)^\dagger$ for all $j$.

\subsubsection{Closed-form composition of translation generators}

Here we provide the closed form expression for powers of $T_{1}$
and $T_{2},$ namely
\begin{align}
T_{2}^{M}c_{\bm{r}}^{\dagger}T_{2}^{-M} & =i^{M}c_{\bm{r}+M\bm{a}_{2}}^{\dagger}\prod_{j=0}^{M-1}e^{i(\bm{r}+j\bm{a}_{2})\cdot\bm{b}_{2}/2}\\
 & =i^{M}c_{\bm{r}+M\bm{a}_{2}}^{\dagger}e^{i\sum_{j=0}^{M-1}\bm{r}\cdot\bm{b}_{2}/2}e^{i\pi\sum_{j=0}^{M-1}j}\\
 & =i^{M}c_{\bm{r}+M\bm{a}_{2}}^{\dagger}e^{iM\bm{r}\cdot\bm{b}_{2}/2}e^{i\pi M(M-1)/2}\\
 & =c_{\bm{r}+M\bm{a}_{2}}^{\dagger}e^{iM\bm{r}\cdot\bm{b}_{2}/2}e^{i\pi M^{2}/2},
\end{align}
and similarly
\begin{align}
T_{1}^{M}c_{\bm{r}}^{\dagger}T_{1}^{-M} & =i^{M}c_{\bm{r}+M\bm{a}_{1}}^{\dagger}\prod_{j=0}^{M-1}e^{i(\bm{r}+j\bm{a}_{2})\cdot(\bm{b}_{1}+\bm{b}_{2})/2}\\
 & =i^{M}c_{\bm{r}+M\bm{a}_{1}}^{\dagger}e^{iM\bm{r}\cdot(\bm{b}_{1}+\bm{b}_{2})/2}e^{i\pi M(M-1)/2}\\
 & =c_{\bm{r}+M\bm{a}_{1}}^{\dagger}e^{iM\bm{r}\cdot(\bm{b}_{1}+\bm{b}_{2})/2}e^{i\pi M^{2}/2}.
\end{align}
Then for two-electron operators, in particular, we obtain:
\begin{align}
T_{\bm{M}}c_{\bm{r}}^{\dagger}c_{\bm{r}'}^{\dagger}T_{\bm{M}}^{\dagger} & =T_{2}^{M_{2}}T_{1}^{M_{1}}c_{\bm{r}}^{\dagger}c_{\bm{r}'}^{\dagger}T_{1}^{-M_{1}}T_{2}^{-M_{2}}\\
 & =e^{i\pi M_{1}^{2}}e^{iM_{1}(\bm{r}+\bm{r}')\cdot(\bm{b}_{1}+\bm{b}_{2})/2}T_{2}^{M_{2}}c_{\bm{r}+M_{1}\bm{a}_{1}}^{\dagger}c_{\bm{r}'+M_{1}\bm{a}_{1}}^{\dagger}T_{2}^{-M_{2}}\\
 & =e^{i\pi(M_{1}^{2}+M_{2}^{2})}e^{iM_{1}(\bm{r}+\bm{r}')\cdot(\bm{b}_{1}+\bm{b}_{2})/2}e^{iM_{2}(\bm{r}+M_{1}\bm{a}_{1})\cdot\bm{b}_{2}/2}e^{iM_{2}(\bm{r}'+M_{1}\bm{a}_{1})\cdot\bm{b}_{2}/2}c_{\bm{r}+\bm{M}}^{\dagger}c_{\bm{r}'+\bm{M}}^{\dagger} \\
 & =e^{i\pi(M_{1}^{2}+M_{2}^{2})}e^{iM_{1}(\bm{r}+\bm{r}')\cdot(\bm{b}_{1}+\bm{b}_{2})/2}e^{iM_{2}\bm{r}\cdot\bm{b}_{2}/2}e^{iM_{2}\bm{r}'\cdot\bm{b}_{2}/2}c_{\bm{r}+\bm{M}}^{\dagger}c_{\bm{r}'+\bm{M}}^{\dagger}.
\end{align}

\subsection{$C_2$ gauge for XC geometries} \label{app:XC_C2gauge}

Fig.~\ref{fig:two_gauges}(b) displays the Hamiltonian hopping amplitudes that we employ for our calculations on the XC4 cylinder analyzed in the main text. In this gauge, the hoppings are real and imaginary. Moreover, the site-centered inversion symmetry is implemented as $C_2 c^\dag_{\bmr} C_2^{-1} = c^\dag_{C_2 \bmr}$.

The lattice vector $\bm{a}_1$ points along the cylinder axis, while $2\bm{a}_2 - \bm{a}_1$ has length $\sqrt{3}$ and points along the circumference. For the XC4 infinite cylinder~\cite{Szasz2020}, the lattice and hoppings are repeated so that the cylinder has infinite length and circumference $2\sqrt{3}$. We also employ the hoppings of Fig.~\ref{fig:two_gauges}(b) in Sec.~\ref{app:DMRG_xis} of the \textit{Supporting Information} below, where we calculate ground state correlation lengths on the XC4 and XC6 infinite cylinders at half filling.

\section{Details of the U(1) Slave Rotor Theory and Mean Field Estimate of the Transition Point} \label{app:U1_rotor_details}

We utilize the slave-rotor approach~\cite{FlorensGeorges2002,Florens2004}, introducing a U(1) rotor variable $e^{i\theta}$ and its conjugate integer-valued ``angular momentum'' $L$ on each site and writing the electron operator as
\begin{align}
    c^\dagger_{i,\sigma} = f^\dagger_{i,\sigma} \,e^{i\theta_i},   
\end{align}
where $f_\sigma$ is a fermionic ``spinon'' that carries the spin index. The Hilbert space redundancy introduced by the rotors is removed by requiring that each site satisfy a constraint:
\begin{equation}
    L_i = \sum_\sigma \left ( f^\dagger_{i,\sigma} f_{i,\sigma} -1/2 \right ).
\label{eq:constraint}
\end{equation}
Note that the singly-filled site is represented by $L_i=0$, while the doublon/empty sites are $L_i=\pm 1$. The Hubbard model can then be rewritten as:
\begin{equation}
H = -\sum_{\langle ij\rangle,\sigma} t_{ij}e^{i(\theta_i-\theta_j)}f^\dagger_{i,\sigma} f_{j,\sigma} +{\rm h.c.} +\frac{U}{2}\sum_{i}L_i^2,
\end{equation}
supplemented with the constraint Eq.~\eqref{eq:constraint} and the commutation relation $\left [ e^{i\theta_i},\,L_j\right ] = -\delta_{ij}e^{i\theta_i}$.

Of course, treating everything exactly is tantamount to solving the original problem, but we can make progress by adopting a mean field approach, satisfying the constraint \textit{on average} and assigning self-consistent expectation values to $\langle f^\dagger_{i,\sigma} f_{j,\sigma}\rangle$ and $\langle e^{i(\theta_i-\theta_j)}\rangle$. This gives us the mean field Hamiltonian:
\begin{eqnarray}
    H_{MF} &=& H_f + H_\theta\\
    H_{f} &=&  -\sum_{\langle ij\rangle,\sigma} t_{ij}\langle e^{i(\theta_i-\theta_j)}\rangle f^\dagger_{i,\sigma} f_{j,\sigma} +{\rm h.c.} \\
H_{\theta} &=&  -\sum_{\langle ij\rangle,\sigma} t_{ij} e^{i(\theta_i-\theta_j)}\langle f^\dagger_{i,\sigma} f_{j,\sigma}\rangle +{\rm h.c.} +\frac{U}2\sum_{i}L_i^2 
    \label{eq:MF_app}
\end{eqnarray}
Note that the constraint Eq.~\eqref{eq:constraint} implies gauge invariance, $\theta_i\rightarrow \theta_i+\epsilon_i,\, f^\dagger_i\rightarrow f^\dagger_i e^{-i\epsilon_i} $ which is broken by the mean field Hamiltonian. This is remedied by incorporating a fluctuating gauge field $a$ on the links, though we shall not do that here.

In the limit of small Hubbard $U$, we expect the rotor fields to condense, $\langle e^{i\theta_i} \rangle = \sqrt{Z_i} \leq 1$, so that the spinons $f$ are identified with the electron up to this prefactor.
This phase possesses off-diagonal long-range order; at the mean-field level, we assume $\langle e^{i(\theta_i-\theta_j)}\rangle \approx \langle e^{i\theta_i}\rangle \langle e^{-i\theta_j}\rangle$. Thus, the gap to the electron excitation is obtained from the dispersion $t_{ij}\sqrt{Z_iZ_j}f^\dagger_if_j$, which explains why the single-electron gap on the IQH side (\textit{i.e.}, the condensed rotor phase) {\em decreases} on increasing $U$.

On the other side of the phase diagram, when $U$ is sufficiently large, we have $\langle e^{i\theta_i} \rangle =0$ and thus the electron excitation is distinct from the spinon. To create an electron, one must both create a spinon $f^\dagger$ {\em and} increase the rotor angular momentum by $\Delta L =1$. Thus, due to the interaction term, this gap tracks $U$ for large $U$. Despite the absence of an expectation value $\langle e^{i\theta_i} \rangle =0$ on this side, there is virtual tunneling of the rotor quanta, so that $\langle e^{i(\theta_i-\theta_j)} \rangle =\beta_{ij}\neq 0$ for neighboring sites~\cite{Song2023}. In fact, this expectation value is expected to scale as $|t|/U$, in the large $U$ limit, implying that the dispersion of spinons in Eq.~\eqref{eq:MF_app} is on the order of $t^2/U$, as is expected of magnetic excitations. On the other hand, if we assume that the expectation values $\beta_{ij}$ are positive, consistent with the fact that the rotors see no net flux (while the electrons and spinons do), then we simply fill up the negative energy states of the band-structure shown in Fig.~1(c) of the main text, albeit with a smaller gap. This gives a non-zero value for the rotor hoppings, $\langle f^\dagger_{i,\sigma} f_{j,\sigma}\rangle = \alpha_{ij}$. We can find this expectation value from the ground state energy of the band Hamiltonian: writing $\alpha_{ij} = -t^*_{ij}\alpha/|t| $, we expect the average over the filled bands $\langle\epsilon^-(k) \rangle$ to determine $\alpha =\frac{|\langle\epsilon^-(k) \rangle|}{2z|t|} $, where $z=6$ is the coordination number.

Now we can determine the transition, which amounts to analyzing the Bose-Hubbard model
\begin{align}
    H_B = -J \sum_{\langle i j\rangle} e^{i(\theta_i-\theta_j)} + {\rm h.c.} +\frac{U}2\sum_i L_i^2.   
\end{align}
A simple mean field theory locates the transition at $U^* = 4zJ$. This is done by taking a simple variational ansatz for each site, $|\psi\rangle = |0\rangle +\frac{\psi}{2}\left ( |+1\rangle +|-1\rangle \right )$ in the $L$ basis, which has the property $\langle \psi |e^{i\theta}|\psi\rangle = \psi$. Then $\langle H_B\rangle = N\psi^2 (-zJ +\frac{U}4)$. The condensate occurs when the coefficient in parentheses first turns negative.

Substituting $J = 2 |t|\alpha$, where the factor of $2$ is for the two spins of $f$, we get:
\begin{equation}
U^* = 4 |\langle\epsilon^-(k) \rangle| \approx 9.60813 \,t
\end{equation}
where we averaged over the dispersion $\epsilon_-(k) = -2t\sqrt{\cos^2 k_1+\cos^2 k_2+\cos^2 (k_1+k_2)}$. This mean field result is to be compared to, and is indeed relatively close to, the iDMRG-obtained transition that occurs at $U^*/t \approx 11-12 $~\cite{Kuhlenkamp2024,Divic2024}.

Additionally, we can argue directly that the gap to the charge-$2e$ Cooper pair excitations vanishes at the transition. To access them one has to include coupling to the internal gauge field $a$. If we focus on the transition and integrate out the fermionic spinons $f$, this gives a Chern-Simons term~\cite{qi2008topological}. Also, the rotor variables are at low energies and couple minimally to the gauge field. In the condensed phase of the rotors, we have the effective Lagrangian:
\begin{align}
    L = \frac1{4U} \left ( \dot{\theta} + a_0\right )^2 - \frac{\rho_s}{2}\left ( \nabla{\theta} + a\right )^2 + \frac2{4\pi} a\wedge da,    
\end{align}
where the superfluid density $\rho_s$ goes to zero at the transition. The vortices of the rotor condensate are the Cooper pairs. The fact that they carry $2\pi$ flux of $da$ implies, via the Chern-Simons term, a gauge charge of $2$. This is screened by two rotor fields, giving the vortex a global $\mathrm{U(1)}_c$ charge of $2$ and no spin. The energy of such vortices vanishes as we approach the transition.

\section{Chirality of Edge States from K-Matrix Formalism}

In this appendix, we provide a more thorough account of the charge response, spin response, and edge states of the IQH, CSL, and superconductor phases within the framework of Chern-Simons K-matrix formalism~\cite{WenZee1992_Kmatrix}. We employ the following convention for the Chern-Simons Lagrangian: 
\begin{equation} \label{eq:Kmatrix_lagrangian}
\mathcal{L}=-\frac{K_{IJ}}{4\pi}\alpha_{I}\wedge d\alpha_{J}+\frac{t_{J}}{2\pi}A\wedge d\alpha_{J}+\frac{(t_{s})_{J}}{2\pi}A_{s}\wedge d\alpha_{J},
\end{equation}
or in components:
\begin{equation}
\mathcal{L}=-\frac{1}{4\pi}\alpha_{I}^{\mu}K_{IJ}\epsilon_{\mu\nu\lambda}\partial^{\nu}\alpha_{J}^{\lambda}+\frac{1}{2\pi}A^{\mu}t_{J}\epsilon_{\mu\nu\lambda}\partial^{\nu}\alpha_{J}^{\lambda}+\frac{1}{2\pi}(A_{s})^{\mu}(t_{s})_{J}\epsilon_{\mu\nu\lambda}\partial^{\nu}\alpha_{J}^{\lambda}.
\end{equation}
Here, $\alpha_{I}$ are dynamical U(1) gauge fields. On the other
hand, $A$ is a classical field that probes the charge
response, while $A_{s}$ probes the spin response. In order for this effective theory
to describe a microscopic system of electrons, it must obey the ``spin/charge
relation'', which is the following requirement on gauge-invariant
\textit{local} operators: those with odd electric charge must have
fermionic statistics, while those with even electric charge have bosonic
statistics~\cite{SeibergWitten_weakcoupling}. Mathematically, this condition reads (see
Section~2.3 of Ref.~\cite{SeibergWitten_weakcoupling}):
\begin{align} \label{eq:spinchargerel}
t_{I}\equiv K_{II}\ (\text{mod }2).
\end{align}
This relation will be satisfied by the K-matrix theories for the IQH,
CSL, and superconductor described below.

On a spatial manifold of genus $g,$ the ground state degeneracy is
given by $|\text{det} K|^{g}$, provided $\text{det} K\neq0$. Furthermore, the
edge chiral central charge is given by the ``signature'' of $K,$ \textit{i.e.},
$c_{-}=$ the number of positive eigenvalues minus the number of negative
eigenvalues~\cite{LuVishwanath2016}.
Provided the K-matrix is invertible, the Hall conductivity is~\cite{WenZee1992_Kmatrix}
\begin{equation}
\sigma_{xy}=\frac{e^{2}}{h}\sum_{IJ}t_{I}(K^{-1})_{IJ}t_{J}.
\end{equation}
and the spin quantum Hall conductivity (\textit{i.e.}, spin current in response
to a Zeeman gradient) is likewise
\begin{equation}
\sigma_{xy}^{s}=\frac{(\hbar/2)^{2}}{h}\sum_{IJ}(t_{s})_{I}(K^{-1})_{IJ}(t_{s})_{J}.
\end{equation}
Moreover, within the K-matrix formalism, each quasi-particle excitation
can be labeled by a $\bm{\ell},$ where $\ell_{I}$ denotes its charge
under each of the dynamical gauge fields. The current of this quasi-particle
is then a source for the dynamical gauge fields and has statistical angle
\begin{equation}
\theta(\bm{\ell})=\pi\sum_{IJ}\ell_{I}(K^{-1})_{IJ}\ell_{J},
\end{equation}
electric charge
\begin{equation}
q(\bm{\ell})=e\sum_{IJ}t_{I}(K^{-1})_{IJ}\ell_{J},
\end{equation}
(where $e$ is the electron charge) and $S_{z}$ spin quantum number
\begin{equation}
q_{s}(\bm{\ell})=\frac{\hbar}{2}\sum_{IJ}(t_{s})_{I}(K^{-1})_{IJ}\ell_{J}.
\end{equation}

The spin-singlet integer quantum Hall phase (with total Chern number
$C=2)$ is described by the data
\begin{equation}
K_{\text{IQH}}=\begin{pmatrix}1 & 0\\
0 & 1
\end{pmatrix},\qquad t_{\text{IQH}}=\begin{pmatrix}1\\
1
\end{pmatrix},\qquad t_{s,\text{IQH}}=\begin{pmatrix}1\\
-1
\end{pmatrix}.
\end{equation}
One immediately verifies that it satisfies the expected properties.
It has a unique gapped ground state, as well as $c_{-}=2,$ $\sigma_{xy}=2\cdot e^{2}/h,$
and $\sigma_{xy}^{s}=2\cdot(\hbar/2)^{2}/h.$ Moreover, no quasi-particles
are fractionalized: since $K_{\text{IQH}}=K_{\text{IQH}}^{-1}=1_{2\times2},$
then $q(\bm{\ell})=(\ell_{1}+\ell_{2})e$ and $q_{s}(\bm{\ell})=(\ell_{1}-\ell_{2})\hbar/2.$
Thus, in units of $e$ and $\hbar/2$ respectively, $q(\bm{\ell})$
and $q_{s}(\bm{\ell})$ are both odd or both even.

The CSL phase is obtained by ``gauging'' the global charge U(1)
of the IQH phase. Specifically, let us promote $A\to\alpha_{3}+A,$
where $\alpha_{3}$ is a third dynamical U(1) gauge field and $A$
is still a classical field. The resulting theory is
described by the data
\begin{equation}
K_{\text{CSL}}=\begin{pmatrix}1 & 0 & -1\\
0 & 1 & -1\\
-1 & -1 & 0
\end{pmatrix},\qquad t_{\text{CSL}}=\begin{pmatrix}1\\
1\\
0
\end{pmatrix},\qquad t_{s,\text{CSL}}=\begin{pmatrix}1\\
-1\\
0
\end{pmatrix}.
\end{equation}
The eigenvalues of $K_{\text{CSL}}$ are $2,1,-1.$ Thus, the ground
state degeneracy is $2^{g}$ and the chiral central charge is $c_{-}=1,$
as expected in the Kalmeyer-Laughlin CSL phase. Moreover, since
\begin{equation}
(K_{\text{CSL}})^{-1}=\frac{1}{2}\begin{pmatrix}1 & -1 & -1\\
-1 & 1 & -1\\
-1 & -1 & -1
\end{pmatrix},
\end{equation}
then this phase has the expected quantized Hall responses, $\sigma_{xy}=0$
and $\sigma_{xy}^{s}=2\cdot(\hbar/2)^{2}/h.$ Within this effective
theory, the spin-up (spin-down) electron excitation is labeled by
the first column $\bm{\ell}_{e\uparrow}$ (second column $\bm{\ell}_{e\downarrow})$
of $K_{\text{CSL}},$ with charges
\begin{equation}
q(\bm{\ell}_{e,\uparrow/\downarrow})=e,\qquad q_{s}(\bm{\ell}_{e,\uparrow/\downarrow})=\pm\hbar/2.
\end{equation}
On the other hand, there is a fractionalized charged semion excitation
labeled by $\bm{\ell}_{c}=(\begin{matrix}0 & 0 & -1\end{matrix})^{\mathsf{T}},$
 which carries $q(\bm{\ell}_{c})=e,q_{s}(\bm{\ell}_{c})=0,$ and a
semionic statistical angle $\theta(\bm{\ell}_{c})=-\pi/2.$ The fractionalized
spinful semion excitations correspond to $\bm{\ell}_{s,\uparrow/\downarrow}=\bm{\ell}_{e,\uparrow/\downarrow}-\bm{\ell}_{c},$
which indeed have $q(\bm{\ell}_{s})=0,q_{s}(\bm{\ell}_{s,\uparrow/\downarrow})=\pm\hbar/2,$
and a semionic statistical angle $\theta(\bm{\ell}_{s})=+\pi/2.$

Now, let us turn to the anyon superconductor and the chirality of
its edge modes. From the main text, recall that the main additional
step was that the gauge field $a=\alpha_{3}$ now has a bIQH response
$-\frac{2}{4\pi}ada$. With the sign convention of Eq.~\eqref{eq:Kmatrix_lagrangian}, adding
this to $K_{\text{CSL}}$ yields
\begin{equation} \label{eq:SC_Kdata}
K_{\text{SC}}=\begin{pmatrix}1 & 0 & -1\\
0 & 1 & -1\\
-1 & -1 & 2
\end{pmatrix},\qquad t_{\text{SC}}=\begin{pmatrix}1\\
1\\
0
\end{pmatrix},\qquad t_{s,\text{SC}}=\begin{pmatrix}1\\
-1\\
0
\end{pmatrix}.
\end{equation}
We find $\det K_{\text{SC}}=0,$ which signals that the gauge fields are not independent and that the system (absent $A$) is gapless. Performing the change of variables
\begin{align}
    \begin{pmatrix}\alpha_{1}\\
    \alpha_{2}\\
    \alpha_{3}
    \end{pmatrix}=\begin{pmatrix}\alpha_{1}'+\alpha_{3}'\\
    \alpha_{2}'+\alpha_{3}'\\
    \alpha_{3}'
    \end{pmatrix}=W\begin{pmatrix}\alpha_{1}'\\
    \alpha_{2}'\\
    \alpha_{3}'
    \end{pmatrix},\qquad W=\begin{pmatrix}1 & 0 & 1\\
    0 & 1 & 1\\
    0 & 0 & 1
    \end{pmatrix},
\end{align}
we obtain the equivalent K-matrix theory
\begin{align} \label{eq:transformed_KSC}
    K_{\text{SC}}'=W^{\mathsf{T}}K_{\text{SC}}W=\begin{pmatrix}1 & 0 & 0\\
0 & 1 & 0\\
0 & 0 & 0
\end{pmatrix},\qquad t'_\text{SC} =W^{\mathsf{T}}t_\text{SC}=\begin{pmatrix}1\\
1\\
2
\end{pmatrix},\qquad t'_{s,\text{SC}}=W^{\mathsf{T}}t_{s,\text{SC}}=\begin{pmatrix}1\\
-1\\
0
\end{pmatrix}.
\end{align}
This theory obeys the spin/charge relation Eq.~\eqref{eq:spinchargerel}, so that $W$ is a valid transformation. Moreover, we see that $\alpha'_3$ has decoupled from the other gauge fields and is governed by its Maxwell dynamics. With $A=0$, this leads to a single bulk gapless mode corresponding to the superfluid Goldstone mode of the spontaneously-broken $\mathrm{U(1)}_c$ symmetry. From the remaining fields, we see that there is no accompanying topological order and that the resulting phase is a chiral superfluid with edge chiral central charge $c_{-}=2$.

Further insight into the topological properties of the superconductor is gained by replacing $A$ with a 2+1D dynamical gauge field~\cite{HANSSON2004497,Moroz2017}. This gaps out the Goldstone mode of the superfluid and gives rise to one of Kitaev's 16-fold way $\mathbb{Z}_2$ topological orders~\cite{Kitaev2006}. For instance, a topologically-trivial superfluid maps to the $m = 0$ member of the 16-fold way.\footnote{We use the symbol $m$ instead of Kitaev's $\nu$ to avoid collision with the filling of the various quantum Hall phases.} In fact, we only expect one of the eight \textit{even} members of the 16-fold way since we will only describe Abelian topological orders via the K-matrix. Making $\alpha'_4 = A$ dynamical in Eq.~\eqref{eq:transformed_KSC} and discarding $A_s$, we obtain
\begin{align}
    K'_\mathrm{gauged\ SC} = \begin{pmatrix}
                       1 & 0 & 0 & -1  \\
                       0 & 1 & 0 & -1  \\
                       0 & 0 & 0 & -2  \\
                       -1 & -1 & -2 & 0 \end{pmatrix},
\end{align}
which has precisely the same topological order as the gauged, weak-pairing $d+id$ superconductors~\cite{Moroz2017}. In particular, it corresponds to the member $m = 4$ of Kitaev's classification~\cite{Kitaev2006}, which has $c_- = m/2 = 2$. However, recall from the main text that our superconductor is far from the weak-pairing limit (evidenced by electron pairing above the insulators), so that its pairing symmetry is not directly linked to its edge chiral central charge~\cite{ReadGreen2000}.

\section{Correlation Lengths at Half-Filling from Cylinder iDMRG} \label{app:DMRG_xis}

In this section, we provide the transfer matrix correlation length data for the ground states at half-filling, for a range of Hubbard interaction strengths $U$ and various system sizes. We consider both the YC-$L_y$ and XC-$2L_y$ cylinder geometries~\cite{Szasz2020}.

Let us first discuss the YC cylinder geometries. In the gauge-invariant language of Sec.~1 of the main text, in the YC case with even $L_y$, we take $T = T_2$ to be a purely circumferential translation with $T^{L_y}=1$. When $L_y$ is odd, we instead thread $\pi/2$ flux so that $T^{L_y}= i^{N_F}$, which means that $T$ can have eigenvalue $-1$ in the charge-$2e$ $(N_F=2)$ sector. In both cases, the fluxes through the rings are such that the system respects particle-hole symmetry.

The XC case proceeds the same way except that we take our circumferential translation to be $T = (T_2)^2 T_1^\dag$, which translates by a distance $\sqrt{3}a$ instead of $a$. For our Hamiltonian hopping amplitudes, we specifically choose those shown in Fig.~\ref{fig:two_gauges}(b).

The dominant correlation length $\xi$ in a given charge sector of the transfer matrix is related to the dominant non-trivial transfer matrix eigenvalue $\lambda$ by~\cite{Zauner2015NJPh}:
\begin{align}
    1/\xi_i = -\log |{\lambda_i}|.
\end{align}
We report $\xi$ in units of cylinder rings. For the YC cylinders, the distance between the cylinder rings is $a\sqrt{3}/2$, whereas it is $a/2$ in the XC case.

\begin{figure*}
    \centering
    \includegraphics[width = 494 pt]{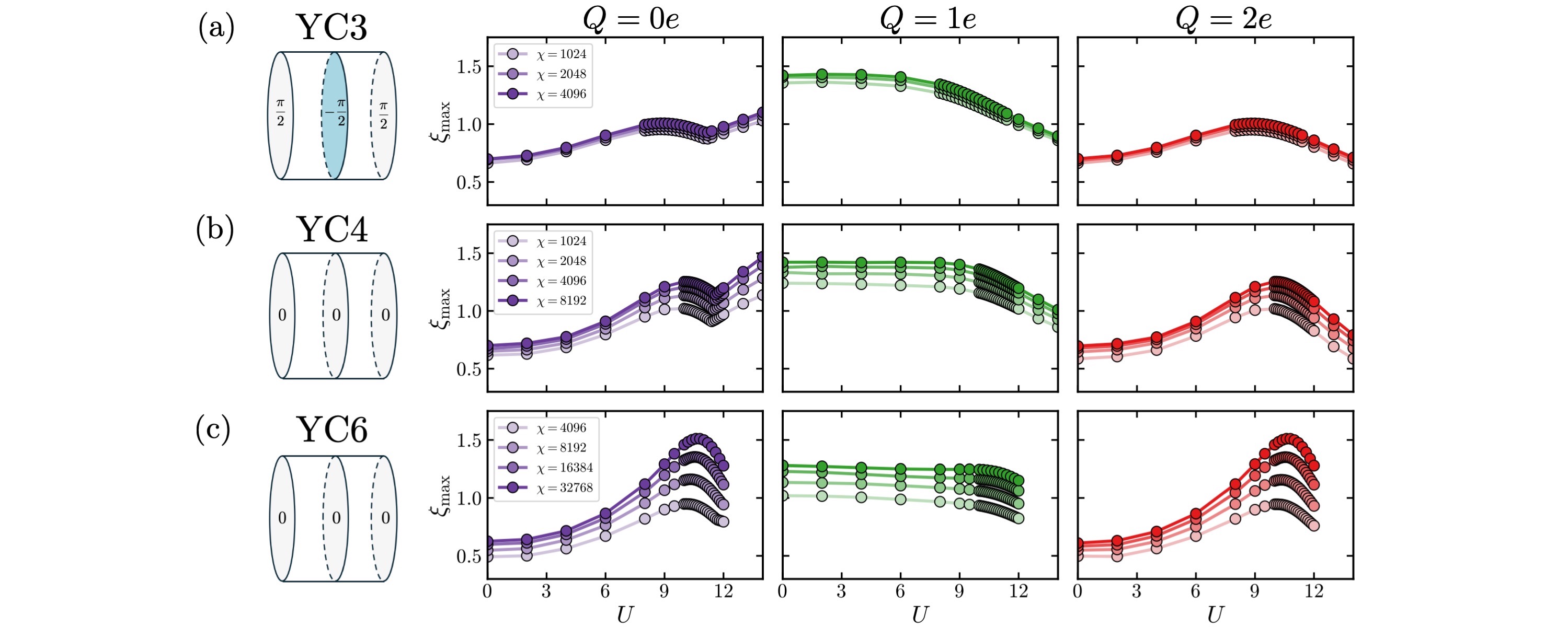}
    \caption{(a) Maximum correlation length of the half-filled ground state transfer matrix on the infinite YC3 cylinder vs. Hubbard $U$ in the charge sectors $Q\in\{0e,1e,2e\}$. The external flux is chosen so that the cylinder rings are pierced by fluxes $\pi/2,-\pi/2,\dots$ which alternate between rings. Bond dimension, increasing with shade light-to-dark, indicated in legend. (b) Same for the YC4 infinite cylinder with zero flux through each ring. (c) Same for YC6 with zero flux through each ring.
    }
    \label{fig:YC_xis}
\end{figure*}

\begin{figure*}
    \centering
    \includegraphics[width = 494 pt]{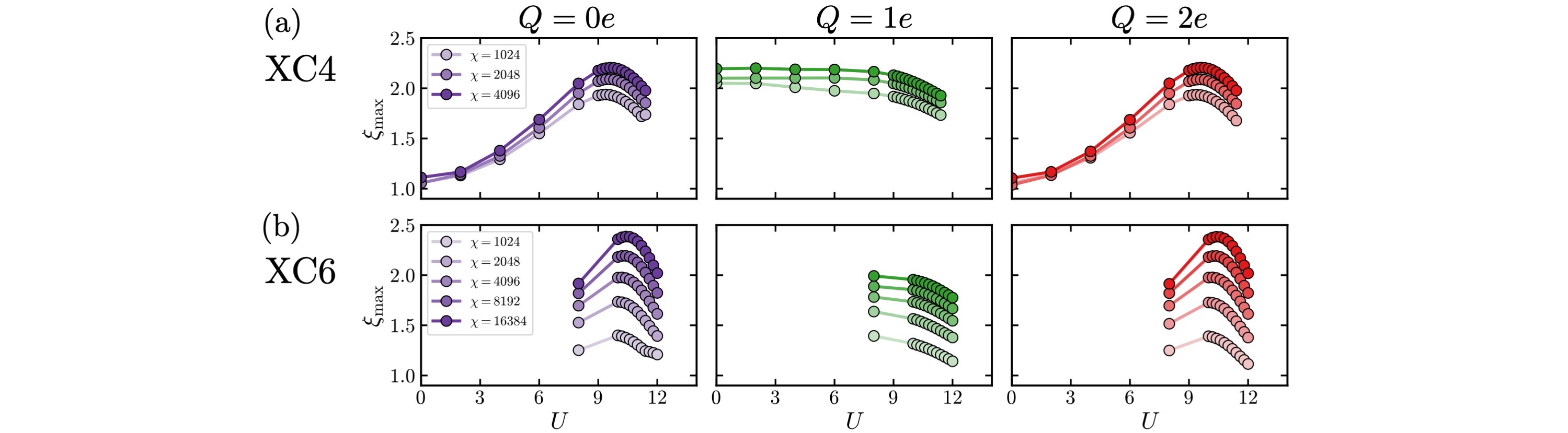}
    \caption{(a) Maximum correlation length of the half-filled ground state transfer matrix on the infinite XC4 cylinder vs. Hubbard $U$ in the charge sectors $Q\in\{0e,1e,2e\}$. Bond dimension, increasing light-to-dark with shade, indicated in legend.
    (b) Same for XC6 cylinder. For both the XC4 and XC6 cylinders, we make the choice of Hamiltonian hopping amplitudes specified in Sec.~\ref{app:XC_C2gauge} of the \textit{Supporting Information}.
    }
    \label{fig:XC_xis}
\end{figure*}

In Fig.~\ref{fig:YC_xis}, we display the maximum correlation lengths for the YC3, YC4 and YC6 cylinders resolved by the electric charge sector of the ground state transfer matrix, namely $Q \in \{0e, 1e, 2e\}$, where we maximize over spin and momentum quantum numbers for simplicity. We note that due to the pseudospin SU(2) symmetry, the charge-$0e$ and charge-$2e$ sectors manifestly have equal correlation lengths, with the exception of the regime $U/t \gtrsim 11$ where the spin-\textit{triplet} charge-$0e$ branch (known to be low-energy from the ED and DMRG excitation calculations in the main text) have quantitatively larger correlation length, at least for the YC3 and YC4 systems. The main numerical feature we point out here, consistent with Ref.~\cite{Divic2024} (though differentiated here since we strictly study the pseudospin-symmetric configurations), is that the correlation length peak and the sharpness of the peak is in the charge-$0e/2e$ sectors is increasing with system size while the charge-$1e$ sector does not see a similar enhancement. The same features are seen in the XC4 vs. XC6 data, shown in Fig.~\ref{fig:XC_xis}.

\section{Spin and Pseudospin SU(2) Formalism}

\subsection{Relation between spin and pseudospin generators}

Here we present an alternative description, complementing the Majorana representation discussed in Sec.~1 of the main text, of how the presence of particle-hole symmetry and Hubbard interactions gives rise to the pseudospin $\mathrm{SU}(2)_c$ symmetry~\cite{Shiroishi_1998}. We begin with the familiar spin rotation generators:
\begin{align}
S^{+}=\sum_{\bm{r}}c_{\bm{r}\uparrow}^{\dagger}c_{\bm{r}\downarrow},\qquad S^{-}=\sum_{\bm{r}}c_{\bm{r}\downarrow}^{\dagger}c_{\bm{r}\uparrow},\qquad S^{z}=\frac{1}{2}\sum_{\bm{r}}(n_{\bm{r}\uparrow}-n_{\bm{r}\downarrow}).
\end{align}
Given a particle hole symmetry $\mathcal{P}c_{\bm{r}\sigma}\mathcal{P}^{-1} = e^{i\alpha_{\bm{r}}}c_{\bm{r}\sigma}^{\dagger}$, which commutes with the Hamiltonian, let us define the following \textit{half-}particle-hole transformation:
\begin{align}
\bar{\mathcal{P}}c_{\bm{r}\uparrow}\bar{\mathcal{P}}^{-1} = e^{i\alpha_{\bm{r}}}c^\dag_{\bm{r}\uparrow}, \qquad \bar{\mathcal{P}}c_{\bm{r}\downarrow}\bar{\mathcal{P}}^{-1} = c_{\bm{r}\downarrow}.
\end{align}
Though the half-PH leaves the spin-rotation-symmetric hopping Hamiltonian invariant, it is not a symmetry of the interaction since it flips the sign of the Hubbard $U$:
\begin{align}
\bar{\mathcal{P}}\left(U\sum_{\bm{r}}(n_{\bm{r}\uparrow}-1/2)(n_{\bm{r}\downarrow}-1/2)\right)\bar{\mathcal{P}}^{-1}=U\sum_{\bm{r}}(1-n_{\bm{r}\uparrow}-1/2)(n_{\bm{r}\downarrow}-1/2)=-U\sum_{\bm{r}}(n_{\bm{r}\uparrow}-1/2)(n_{\bm{r}\downarrow}-1/2).
\end{align}
Since $H_{-U}$ also commutes with the spin generators, then $\bar{\mathcal{P}}S^{\alpha}\bar{\mathcal{P}}^{-1}$
are symmetries of $H_{+U}.$ In particular, note that
\begin{align}
\eta^{+}\equiv\bar{\mathcal{P}}S^{+}\bar{\mathcal{P}}^{-1}=\sum_{\bm{r}}e^{-i\alpha_{\bm{r}}}c_{\bm{r}\uparrow}c_{\bm{r}\downarrow},
\end{align}
while
\begin{align}
\eta^{z}\equiv\bar{\mathcal{P}}S^{z}\bar{\mathcal{P}}^{-1}=\frac{1}{2}\sum_{\bm{r}}(1-n_{\bm{r}\uparrow}-n_{\bm{r}\downarrow})
\end{align}
is related to the total charge density. While the resulting symmetries seem unnatural (in that $\eta^z$ includes a minus sign relative to the charge density and $\eta^+$ \textit{lowers} the total charge by $2e$), we will show later that this convention is the most natural for implementing spin and pseudospin rotations simultaneously as left and right actions, respectively, in a way that is consistent with $\bar{\mcP}$ relating the generators, as in Eqs.~(6-8) of Ref.~\cite{Hermele2007}. Also, note that one can simply take $e^{i\alpha_{\bm{r}}}=1$ when the hoppings are all imaginary, so that:
\begin{align} \label{eq:simple_PH}
    \mathcal{P}c_{\bm{r}\sigma}\mathcal{P}^{-1} = c_{\bm{r}\sigma}^{\dagger}.
\end{align}

\subsection{Fermionic matrix structure in $\mathrm{SU}(2)_s\times\mathrm{SU}(2)_c$ formalism} \label{sec:matrix_formalism}

In order to get an analytic handle on how various $0e$ and $\pm 2e$ objects transform into one another under the pseudo SU(2) symmetry, it is useful to establish several identities. In particular, here we motivate defining the $2\times2$ fermion construction
where spin (pseudospin) rotations act on the left (right). We know that
the column vector of fermion annihilation operators transforms as
a spin doublet: defining
\begin{align}
C_{\bm{r}}=\begin{pmatrix}c_{\bm{r}\uparrow}\\
c_{\bm{r}\downarrow}
\end{pmatrix},
\end{align}
then $[S_{a},c_{s}] = -\frac{(\sigma_{a})_{ss'}}{2}c_{s'}$, so that
\begin{align}
[\bm{S},C_{\bm{r}}]=-\frac{\bm{\sigma}}{2}C_{\bm{r}},\qquad[\bm{S},C_{\bm{r}}^{\dagger}]=C_{\bm{r}}^{\dagger}\frac{\bm{\sigma}}{2}.
\end{align}
Exponentiating these relations, we get
\begin{align}
e^{i\bm{\theta}\cdot\bm{S}}C_{\bm{r}}e^{-i\bm{\theta}\cdot\bm{S}}=e^{-i\bm{\theta}\cdot\bm{\sigma}/2}C_{\bm{r}}.
\end{align}

Now it's easy to obtain the analogous relation for the pseudospin SU(2)
generators via the mapping $\bm{\eta}=\bar{\mathcal{P}}\bm{S}\bar{\mathcal{P}}^{\dagger}.$
We simply conjugate both sides by $\bar{\mathcal{P}}$ and use the
fact that $\bar{\mathcal{P}}C_{\bm{r}}\bar{\mathcal{P}}^{\dagger}=\begin{pmatrix}c_{\bm{r}\uparrow}^{\dagger}\\
c_{\bm{r}\downarrow}
\end{pmatrix}.$ For the infinitesimal relations, we obtain $[\bm{\eta},\begin{pmatrix}c_{\bm{r}\uparrow}^{\dagger}\\
c_{\bm{r}\downarrow}
\end{pmatrix}]=-\frac{\bm{\sigma}}{2}\begin{pmatrix}c_{\bm{r}\uparrow}^{\dagger}\\
c_{\bm{r}\downarrow}
\end{pmatrix},$ or by taking the Hermitian conjugate,
\begin{align}
\left[\bm{\eta},\begin{pmatrix}c_{\bm{r}\uparrow} & c_{\bm{r}\downarrow}^{\dagger}\end{pmatrix}\right]=\begin{pmatrix}c_{\bm{r}\uparrow} & c_{\bm{r}\downarrow}^{\dagger}\end{pmatrix}\frac{\bm{\tau}}{2}.
\end{align}
(we have relabeled the Pauli matrices as $\bm{\tau}$ when
they act to the right as pseudospin rotations). Exponentiating gives

\begin{align}
e^{i\bm{\theta}\cdot\bm{\eta}}\begin{pmatrix}c_{\bm{r}\uparrow} & c_{\bm{r}\downarrow}^{\dagger}\end{pmatrix}e^{-i\bm{\theta}\cdot\bm{\eta}}=\begin{pmatrix}c_{\bm{r}\uparrow} & c_{\bm{r}\downarrow}^{\dagger}\end{pmatrix}e^{i\bm{\theta}\cdot\bm{\tau}/2}.
\end{align}
Thus $\begin{pmatrix}c_{\bm{r}\uparrow} & c_{\bm{r}\downarrow}^{\dagger}\end{pmatrix}$ should be a row of the $2\times2$ fermion matrix. It turns out that completing this matrix with $-c_{\bm{r}\uparrow}^{\dagger}$ as the bottom right element makes both columns (rows) transform as spin (charge) doublets. Namely,
\begin{align}
\Psi_{\bm{r}}=\begin{pmatrix}c_{\bm{r}\uparrow} & c_{\bm{r}\downarrow}^{\dagger}\\
c_{\bm{r}\downarrow} & -c_{\bm{r}\uparrow}^{\dagger}
\end{pmatrix} = \begin{pmatrix}C_{\bmr} & i\sigma_y C_{\bmr}^* \end{pmatrix}
\end{align}
satisfies
\begin{align}
e^{i\bm{\theta}\cdot\bm{S}} \Psi_{\bm{r}}e^{-i\bm{\theta}\cdot\bm{S}}=e^{-i\bm{\theta}\cdot\bm{\sigma}/2} \Psi_{\bm{r}}
\end{align}
and, on account of the relation $\bar{\mathcal{P}} C_{\bm{r}}\bar{\mathcal{P}}^{\dagger}= C_{\bm{r}}^{\dagger},$
also satisfies
\begin{align}
e^{i\bm{\theta}\cdot\bm{\eta}} \Psi_{\bm{r}}e^{-i\bm{\theta}\cdot\bm{\eta}}= \Psi_{\bm{r}}e^{i\bm{\theta}\cdot\bm{\tau}/2}.
\end{align}
We also remark that the spin and charge generators can be neatly written in terms of this fermionic matrix. First:
\begin{align}
\bm{S}_{\bmr}=\frac{1}{4}\text{tr}\left( \Psi_{\bm{r}}^{\dagger}\bm{\sigma} \Psi_{\bm{r}}\right),
\end{align}
which naturally transforms as a vector under spin rotations. Using the half-PH change of basis, we directly obtain
\begin{align}
\bm{\eta}_{\bmr}=\bar{\mathcal{P}}\bm{S}\bar{\mathcal{P}}^{\dagger}=\frac{1}{4}\text{tr}\left(\Psi_{\bm{r}}\bm{\tau} \Psi_{\bm{r}}^{\dagger}\right),
\end{align}
where we have once again trivially rewritten the charge Pauli matrices as $\bm{\tau}.$

Finally, we note a useful implementation of the Hermitian conjugate:
\begin{align} \label{eq:conj_trick}
 \Psi_{\bm{r}}^{\dagger}=\begin{pmatrix}c_{\bm{r}\uparrow}^{\dagger} & c_{\bm{r}\downarrow}^{\dagger}\\
c_{\bm{r}\downarrow} & -c_{\bm{r}\uparrow}
\end{pmatrix}=-\sigma_{y}\begin{pmatrix}c_{\bm{r}\uparrow} & c_{\bm{r}\downarrow}\\
c_{\bm{r}\downarrow}^{\dagger} & -c_{\bm{r}\uparrow}^{\dagger}
\end{pmatrix}\sigma_{y}=-\tau_{y} \Psi_{\bm{r}}^{\mathsf{T}}\sigma_{y}.
\end{align}
This can be useful when contracting multiple $\Psi_{\bmr}$ under a trace, but one has to be careful about extra minus signs from fermion anti-commutation.

\subsection{Hofstadter-Hubbard Hamiltonian}

Suppose we are given a hopping model $h = -\sum_{\bmr,\bmr'} t_{\bmr\bmr'} c^\dag_{\bmr} c_{\bmr'}$ where $\bmr,\bmr'$ each run over all the sites in the triangular lattice. As usual, Hermiticity requires that
\begin{align}\label{eq:hermit_condition}
t_{\bm{r}\bm{r}'}=t_{\bm{r}'\bm{r}}^{*}.
\end{align}
If we further suppose that the \textit{hoppings are all imaginary}, then
\begin{align}\label{eq:imag_condition}
t_{\bm{r}\bm{r}'}=-t_{\bm{r}'\bm{r}}.
\end{align}
In particular, this means the Hamiltonian has the particle-hole symmetry
$\mathcal{P}c_{\bm{r}}\mathcal{P}^{-1}=c_{\bm{r}}^{\dagger}$, like our $\Phi_{\text{\ensuremath{\triangle}}}=\pi/2$ Hofstadter model on the triangular lattice. In the general case, we claim that the hopping
Hamiltonian can be written as
\begin{align}
h=-\sum_{\bm{r}\leftarrow\bm{r}'} t_{\bm{r}\bm{r}'}\text{tr}\left( \Psi_{\bm{r}}^{\dagger} \Psi_{\bm{r}'}\right),
\end{align}
where the sum is over \textit{oriented} pairs of sites, which we denote by $\bm{r}\leftarrow\bm{r}'$, whose orientation we fix so that $t_{\bm{r}\bm{r}'} = +it$ (with $t > 0$) which gives
\begin{align} \label{eq:simplified_hop}
h=-it\sum_{\bm{r}\leftarrow\bm{r}'} \text{tr}\left( \Psi_{\bm{r}}^{\dagger} \Psi_{\bm{r}'}\right).
\end{align}

The Hamiltonian is already Hermitian in this form. To see this, we explicitly compute its Hermitian conjugate:
\begin{align}
    h^{\dagger}&=-\sum_{\bm{r}\leftarrow\bm{r}'}t_{\bm{r}\bm{r}'}^{*}\text{tr}\left(\Psi_{\bm{r}'}^{\dagger}\Psi_{\bm{r}}\right)=+\sum_{\bm{r}\leftarrow\bm{r}'}t_{\bm{r}\bm{r}'}\text{tr}\left(\left(-\tau_{y}\Psi_{\bm{r}'}^{\mathsf{T}}\sigma_{y}\right)\left(-\sigma_{y}(\Psi_{\bm{r}}^{\dagger})^{\mathsf{T}}\tau_{y}\right)\right),
\end{align}
where we used Eq.~\eqref{eq:conj_trick}. After the Pauli matrices and minus signs cancel, we anti-commute the two operators (they live on different sites) and confirm that
\begin{align}
    h^{\dagger}= - \sum_{\bm{r}\leftarrow\bm{r}'} t_{\bm{r}\bm{r}'}\text{tr}\left((\Psi_{\bm{r}}^{\dagger})^{\mathsf{T}}\Psi_{\bm{r}'}^{\mathsf{T}}\right)=-\sum_{\bm{r}\leftarrow\bm{r}'}t_{\bm{r}\bm{r}'}\text{tr}\left(\Psi_{\bm{r}}^{\dagger}\Psi_{\bm{r}'}\right)=h.
\end{align}

\noindent We note that it is manifestly spin- and pseudospin-rotation-invariant due to the trace. To verify that $h$ is indeed the familiar hopping Hamiltonian, we explicitly expand out the expression:
\begin{align}
h &= -\sum_{\bm{r}\leftarrow\bm{r}'}t_{\bm{r}\bm{r}'}\text{tr}\left(\begin{pmatrix}c_{\bm{r}\uparrow}^{\dagger} & c_{\bm{r}\downarrow}^{\dagger}\\
c_{\bm{r}\downarrow} & -c_{\bm{r}\uparrow}
\end{pmatrix}\begin{pmatrix}c_{\bm{r}'\uparrow} & c_{\bm{r}'\downarrow}^{\dagger}\\
c_{\bm{r}'\downarrow} & -c_{\bm{r}'\uparrow}^{\dagger}
\end{pmatrix}\right)\\
 & = -\sum_{\bm{r}\leftarrow\bm{r}'} (t_{\bm{r}\bm{r}'}c_{\bm{r}s}^{\dagger}c_{\bm{r}'s}+t_{\bm{r}\bm{r}'}c_{\bm{r}s}c_{\bm{r}'s}^{\dagger})\\
 & = -\sum_{\bm{r}\leftarrow\bm{r}'} (t_{\bm{r}\bm{r}'}c_{\bm{r}s}^{\dagger}c_{\bm{r}'s}+t_{\bm{r}\bm{r}'}^{*}c_{\bm{r}'s}^{\dagger}c_{\bm{r}s})
 \\
 & = -\sum_{\bm{r}\leftarrow\bm{r}'} (t_{\bm{r}\bm{r}'}c_{\bm{r}s}^{\dagger}c_{\bm{r}'s}+h.c.),
\end{align}
where we again used the fact that $\bm{r}\neq\bm{r}',$ and for the last equality we crucially used the imaginary condition, Eq.~\eqref{eq:imag_condition}.

As for the Hubbard interaction, we reference Ref.~\cite{Hermele2007}:
\begin{align}
H_{U}=\frac{2U}{3}\sum_{\bm{r}}\left(\eta_{\bm{r}}^{2}-\frac{3}{8}\right),
\end{align}
which is clearly also invariant under both spin (because $[\bm{\eta},\bm{S}]=0)$
and pseudospin rotations.

\section{SU(2) Slave-Rotor Theory of IQH-CSL transition} \label{sec:SU2_slaverotor}

The presence of both $\mathrm{SU}(2)_{s}$ and pseudospin symmetries $\mathrm{SU}(2)_{c}$ enables an SU(2) slave-rotor analysis of the IQH, CSL, and their critical point. The construction was introduced by Hermele in Ref.~\cite{Hermele2007} in the context of Hubbard models on bipartite lattices, but applies in the present context with almost no modification. The main difference is that since the triangular lattice has no natural sublattice, we instead work in the Imaginary C6 gauge (with all imaginary hoppings), where the particle-hole acts as in Eq.~\eqref{eq:simple_PH} above. This means we don't need to keep track of a sublattice-dependent sign, \textit{i.e.}, Hermele's $\epsilon_{A/B} = \pm 1$.

\subsection{Physical Hilbert space from parton constraint}

We begin by decomposing the electron operators into spinons and rotors:
\begin{align}
    \Psi_{\bm{r}}=F_{\bm{r}}Z_{\bm{r}},\qquad F_{\bm{r}}=\begin{pmatrix}f_{\bm{r}\uparrow} & f_{\bm{r}\downarrow}^{\dagger}\\
f_{\bm{r}\downarrow} & -f_{\bm{r}\uparrow}^{\dagger}
\end{pmatrix},\qquad Z_{\bm{r}}=\begin{pmatrix}z_{\bm{r}1} & z_{\bm{r}2}\\
-z_{\bm{r}2}^{\dagger} & z_{\bm{r}1}^{\dagger}
\end{pmatrix}.
\end{align}
The spinon operators are arranged like the electron operators, with spin acting by SU(2) rotations from the left, whereas gauge rotations act from the right. The rotor $Z_{\bm{r}}$ is an SU(2) matrix of operators (meaning $z_{\bm{r}1}^{\dagger}z_{\bm{r}1}+z_{\bm{r}2}^{\dagger}z_{\bm{r}2}=1$), with gauge rotations acting from the left and pseudospin rotations acting from the right.
The enlarged slave-rotor Hilbert space at each site is a product of that of the spin-1/2 spinons and that of the SU(2) rotor. Since the slave-rotor Hilbert space is much larger than the physical one, we have to specify which of its states are physical. Following Ref.~\cite{Hermele2007}, one projects to the subspace of gauge singlets, $\bm{J}_{G}(\bm{r})=0$, where:
\begin{align}
\bm{J}_{G}(\bm{r})=\frac{1}{4}\text{tr}\left(F_{\bm{r}}\bm{\mu}F_{\bm{r}}^{\dagger}\right)+\frac{1}{4}\text{tr}\left(Z_{\bm{r}}^{\dagger}\bm{\mu}Z_{\bm{r}}\right),
\end{align}
so that
\begin{align}
e^{i\bm{\alpha}\cdot\bm{J}_{G}(\bm{r})}Z_{\bm{r}}e^{-i\bm{\alpha}\cdot\bm{J}_{G}(\bm{r})} & =e^{-i\bm{\alpha}\cdot\bm{\mu}/2}Z_{\bm{r}},\\
e^{i\bm{\alpha}\cdot\bm{J}_{G}(\bm{r})}F_{\bm{r}}e^{-i\bm{\alpha}\cdot\bm{J}_{G}(\bm{r})} & =F_{\bm{r}}e^{i\bm{\alpha}\cdot\bm{\mu}/2}.
\end{align}
We have denoted the Pauli matrices acting in the gauge space by $\bm{\mu}$, reserving $\bm{\sigma}$ for the spin space and $\bm{\tau}$ for the pseudospin space, as in Sec.~\ref{sec:matrix_formalism} above.

Note that the $F$-$Z$ Hamiltonian---obtained by writing the
Hamiltonian in terms of partons---is a gauge-singlet operator.
It therefore acts within the $\bm{J}_{G}(\bm{r})=0$ subspace and
has the same matrix elements as the electronic Hamiltonian. To construct this subspace, which is four-dimensional at each site, we start with the slave-particle vacuum:
\begin{align}
|0\rangle_{\text{sp}}=|0\rangle_{f}\otimes|0\rangle_{\text{rot}},
\end{align}
where $|0\rangle_{f}$ is the spinon vacuum and $|0\rangle_{\text{rot}}=|\ell_{C}=0,\ell_{G}=0,m_{C}=0,m_{G}=0\rangle$
is the unique rotationally-invariant state in the rotor Hilbert space ($C$ stands for ``charge'' pseudospin and $G$ stands for gauge);
recall from Ref.~\cite{Hermele2007} that only $\ell_{G}=\ell_{C}$ states are allowed
to begin with. In particular, $\bm{J}_{G}(\bm{r})|0\rangle_{\text{rot}}=0$
by definition. Then it can be shown that while $|0\rangle_\mathrm{sp}$ is not a physical state, $f_{\bm{r}\uparrow}^{\dagger}|0\rangle_{\text{sp}}$ is physical, \textit{i.e.}, it is a gauge-singlet. Similarly, $f_{\bm{r}\downarrow}^{\dagger}|0\rangle_{\text{sp}}$
is a physical state. We can then construct the fully-filled and empty
states by applying electronic operators:
\begin{align}
c_{\bm{r}\uparrow}^{\dagger}f_{\bm{r}\downarrow}^{\dagger}|0\rangle_{\text{sp}}=(f_{\bm{r}\uparrow}^{\dagger}z_{\bm{r}1}^{\dagger}-f_{\bm{r}\downarrow}z_{\bm{r}2})f_{\bm{r}\downarrow}^{\dagger}|0\rangle_{\text{sp}}=(z_{\bm{r}1}^{\dagger}f_{\bm{r}\uparrow}^{\dagger}f_{\bm{r}\downarrow}^{\dagger}-z_{\bm{r}2})|0\rangle_{\text{sp}},
\end{align}
and
\begin{align}
c_{\bm{r}\uparrow}f_{\bm{r}\uparrow}^{\dagger}|0\rangle_{\text{sp}}=(f_{\bm{r}\uparrow}z_{\bm{r}1}-f_{\bm{r}\downarrow}^{\dagger}z_{\bm{r}2}^{\dagger})f_{\bm{r}\uparrow}^{\dagger}|0\rangle_{\text{sp}}=(z_{\bm{r}1}+z_{\bm{r}2}^{\dagger}f_{\bm{r}\uparrow}^{\dagger}f_{\bm{r}\downarrow}^{\dagger})|0\rangle_{\text{sp}}.
\end{align}
Since the electron operators are invariant under gauge transformations,
then both of these states are gauge singlets. We have therefore generated
the full four-dimensional physical on-site Hilbert space.

\subsection{Functional integral representation}

Following the path integral construction outlined in Ref.~\cite{Hermele2007}, the electronic Hofstadter-Hubbard Hamiltonian can now be written in the slave-rotor representation:
\begin{align}
    H_\text{HH} = -\sum_{\bmr,\bmr'} \text{tr}\left[(F_{\bmr} Z_{\bmr})^\dag (F_{\bmr'} Z_{\bmr'}) \right] + \frac{2U}{3}\sum_{\bm{r}}\eta_{\bm{r}}^{2},
\end{align}
which must be accompanied by the aforementioned ``gauge-singlet'' constraint
\begin{align}\label{eq:gauge_constraint}
    \bm{J}_{\bm{G}}(\bmr) = 0.    
\end{align}
Working at zero temperature, we introduce imaginary time $\tau \in (-\infty,+\infty)$ and write the functional integral:
\begin{align}
    \mcZ = \int \mcD Z \mcD F \mcD \bma_\tau e^{-S},\qquad S=S_Z +S_F + S_t,
\end{align}
where
\begin{align}
    S_Z &= \frac{3}{4U} \sum_{\bmr}\int d\tau\ \text{tr}\left[Z^\dag_{\bmr}\left(\overleftarrow{\partial}_{\tau} - \frac{i\bma_\tau\cdot\bm{\mu}}{2} \right) \left({\partial}_{\tau} + \frac{i\bma_\tau\cdot\bm{\mu}}{2} \right) Z_{\bmr} \right],  \\
    S_F &= \frac{1}{2} \sum_{\bmr}\int d\tau\ \text{tr}\left[ F_{\bmr} \left({\partial}_{\tau} + \frac{i\bma_\tau\cdot\bm{\mu}}{2} \right) F_{\bmr}^\dag \right], \\
    S_t &= - it \int d\tau \sum_{\bmr\leftarrow\bmr'} \text{tr}\left[(F_{\bmr} Z_{\bmr})^\dag (F_{\bmr'} Z_{\bmr'}) \right]. \label{eq:SU2_hop}
\end{align}
Here and below, $\overleftarrow{\partial}$ denotes differentiation acting to the left, and the gauge field $\bma_\tau$ has been introduced to enforce the on-site gauge constraint (Eq.~\eqref{eq:gauge_constraint}).

So far everything is exact and therefore intractable, so we pass to a mean-field approximation. The first step is to replace the constraint $Z_{\bmr}(\tau) \in\mathrm{SU(2)}$ with an equivalent Lagrange multiplier term:
\begin{align}
    S_\lambda = i\int d\tau\sum_{\bmr} \lambda_{\bmr}(\tau) \left[ \frac{1}{2} \text{tr}(Z^\dag_{\bmr} Z_{\bmr}) - 1 \right].
\end{align}
The next step is to decouple the hopping term into separate spinon
and rotor terms. This is done by introducing a complex Hubbard--Stratonovich
field via the identity~\cite{LeeLee2005,Hermele2007}:
\[
e^{\epsilon\alpha_{\bm{r}\bm{r}'}\beta_{\bm{r}\bm{r}'}}=\frac{\epsilon}{\pi}\int d\eta_{\bm{r}\bm{r}'}d\eta_{\bm{r}\bm{r}'}^{*}\ e^{-\epsilon(|\eta_{\bm{r}\bm{r}'}|^{2}-\eta_{\bm{r}\bm{r}'}\alpha_{\bm{r}\bm{r}'}-\eta_{\bm{r}\bm{r}'}^{*}\beta_{\bm{r}\bm{r}'})}.
\]
Since we have fixed the bond orientation for the hopping in Eq.~\eqref{eq:SU2_hop}
above, and since $S_{t}=t\int d\tau\sum_{\bm{r}\leftarrow\bm{r}'}(-Z_{\bm{r}'}Z_{\bm{r}}^{\dagger})_{ab}(iF_{\bm{r}}^{\dagger}F_{\bm{r}'})_{ba},$
then we may assign $\epsilon=t\Delta\tau>0$ and for each $a,b$ assign
$\alpha_{\bm{r}\bm{r}'}^{ab}=-(Z_{\bm{r}'}Z_{\bm{r}}^{\dagger})_{ab}$
and $\beta_{\bm{r}\bm{r}'}=(iF_{\bm{r}}^{\dagger}F_{\bm{r}'})_{ba},$
which yields:
\begin{align*}
S_{\eta} & =t\int d\tau\sum_{\bm{r}\leftarrow\bm{r}'}\text{tr}\left[(\eta_{\bm{r}\bm{r}'})^{\dagger}\eta_{\bm{r}\bm{r}'}\right],\\
S_{tZ} & =-t\int d\tau\sum_{\bm{r}\leftarrow\bm{r}'}\text{tr}\left[Z_{\bm{r}}^{\dagger}\eta_{\bm{r}\bm{r}'}Z_{\bm{r}'}\right],\\
S_{tF} & =it\int d\tau\sum_{\bm{r}\leftarrow\bm{r}'}\text{tr}\left[F_{\bm{r}'}(\eta_{\bm{r}\bm{r}'})^{\dagger}F_{\bm{r}}^{\dagger}\right].
\end{align*}
One obtains a real free energy at the saddle point provided that $\eta_{\bm{r}\bm{r}'}=\chi_{\bm{r}\bm{r}'}U_{\bm{r}\bm{r}'}$
is a real number times an SU(2) matrix, as can be shown using the identity
Eq.~\eqref{eq:conj_trick} above. In particular,
two of the saddle point equations read $(\eta_{\bm{r}\bm{r}'})^{\dagger}=\langle Z_{\bm{r}'}Z_{\bm{r}}^{\dagger}\rangle$
and $\eta_{\bm{r}\bm{r}'}=(t_{\bm{r}\bm{r}'}/|t_{\bm{r}\bm{r}'}|)\langle F_{\bm{r}}^{\dagger}F_{\bm{r}'}\rangle=i\langle F_{\bm{r}}^{\dagger}F_{\bm{r}'}\rangle.$
We take an ansatz where $\langle F_{\bm{r}}^{\dagger}F_{\bm{r}'}\rangle$
is proportional to $\langle\Psi_{\bm{r}}^{\dagger}\Psi_{\bm{r}'}\rangle$
in the $U=0$ IQH phase, producing a solution $\eta_{\bm{r}\bm{r}'}^{*}$
(at $\bm{a}_{\tau}=0).$

Fluctuations about this solution are parameterized by an SU(2) gauge-field
on the links $\bm{r}\leftarrow\bm{r}'$ and by a fluctuating $\bm{a}_{\tau}$ component.
Integrating out the spinons while retaining these gauge field degrees
of freedom, then in the continuum limit we obtain at leading order
in the gauge fields~\cite{Wen1991}:
\[
S_{\text{CS}}=\frac{i}{4\pi}\int d^{3}x\ \epsilon^{\mu\nu\rho}\text{tr}\left(a_{\mu}\partial_{\nu}a_{\rho} + \frac{2}{3}a_{\mu}a_{\nu}a_{\rho}\right),
\]
which is an SU(2) Chern-Simons term at level 1~\cite{WittenJonespoly1989}.

\subsection{Low-energy $\mathrm{SU(2)}_{1}$ Higgs-Chern-Simons theory}

To describe the IQH-CSL transition, we pass to a continuum description
of the rotor bosons. We begin with the mean-field action
\begin{align}
S_{\text{MF}}^{Z}=-\sum_{\tau,\bm{r}}\left[\text{tr}Z_{x+\epsilon z}^{\dagger}Z_{x}+h.c.\right]-\sum_{\tau}\sum_{\langle\bm{r},\bm{r}'\rangle}\left[\text{tr}Z_{\bm{r},\tau}^{\dagger}Z_{\bm{r}',\tau}+h.c.\right]+\frac{r+r_{c0}}{2}\sum_{\tau,\bm{r}}\left[\text{tr}Z_{x}^{\dagger}Z_{x}\right],
\end{align}
where $r_{c0}$ is chosen so that the lowest boson mode goes gapless
as $r\to0^{+}.$ Following Ref.~\cite{Hermele2007}, we take a spacetime lattice of triangular lattice sites separated by $\epsilon$ increments in imaginary time. For $r>0,$ diagonalizing the above spacetime-translation-invariant
Hamiltonian gives a unique low-energy mode at $\bm{q}=0,$ consistent
with the rotors experiencing no net flux. We write $\Theta(x)\sim Z_{\bm{r},t}$
with $x$ now a continuous spacetime coordinate, which yields:
\begin{align}
\mathcal{L}_{\text{0}}^{\Theta}=\frac{1}{2}\text{tr}(\Theta^{\dagger}\overleftarrow{\partial}_{\nu}\partial_{\nu}\Theta)+\frac{r}{2}\text{tr}(\Theta^{\dagger}\Theta),
\end{align}
where $\overleftarrow{\partial}$ denotes differentiation to the left.
Note that demanding invariance under global SU(2) gauge transformations and pseudospin rotations precludes a linear time-derivative term $\sim\text{tr}(\Theta^{\dagger}\partial_{\tau}\Theta).$ This entirely fixes the
form at $\mathcal{L}_{\text{0}}^{\Theta}$ quadratic order. Now we
add the gauge field fluctuations by promoting the derivative to the
covariant derivative
\begin{align}
\partial_{\nu}\to\partial_{\nu}+\frac{i\bm{a}_{\nu}\cdot\bm{\mu}}{2},
\end{align}
giving
\begin{align} \label{eq:nonabelian_kinetic}
\mathcal{L}_{\text{kin}}^{\Theta}=\frac{1}{2}\text{tr}\left[\Theta^{\dagger}\left(\overleftarrow{\partial}_{\mu}-\frac{ia_{\mu}^{j}\sigma^{i}}{2}\right)\left(\overrightarrow{\partial}_{\mu}+\frac{ia_{\mu}^{j}\sigma^{i}}{2}\right)\Theta\right],
\end{align}
where $\bm{\mu}$ is again the Pauli matrix vector in the gauge space.
Reintroducing the $\mathrm{SU(2)}{}_{1}$ Chern-Simons term from integrating
out the spinons, and the symmetry-allowed quartic term, the full SU(2)
Higgs-Chern-Simons theory is:
\begin{align}
\mathcal{L}_{\text{eff}}=\mathcal{L}_{\text{CS}}+\mathcal{L}_{\text{kin}}^{\Theta}+\frac{1}{2}\text{tr}(\Theta^{\dagger}\Theta)+\frac{\lambda}{4}\left[\text{tr}(\Theta^{\dagger}\Theta)\right]{}^{2}+\dots
\end{align}

\subsection{Conserved current}

As spin excitations are gapped across the IQH-CSL transition, and neither the spin nor pseudospin symmetries are broken, then the closing of the charge gap is associated with gapless bosonic modes that transform as spin-singlet pseudospin-triplets. We now compute the conserved Noether current associated with the pseudospin rotation $\mathrm{SU}(2)_c$ symmetry. The important Lagrangian term for the current is the coupling $\mathcal{L}_{\text{kin}}^{\Theta}$ between the low-energy bosonic field $\Theta$ and the SU(2) gauge field in Eq.~\eqref{eq:nonabelian_kinetic} above. Under the pseudospin SU(2) rotation
\begin{align}
\Theta'=\Theta e^{i\bm{\alpha}\cdot\bm{\tau}/2}\approx\Theta+\Theta\frac{i\bm{\alpha}\cdot\bm{\tau}}{2},
\end{align}
this term transforms to
\begin{align}
(\mathcal{L}_{\text{kin}}^{\Theta})' & =\frac{1}{2}\text{tr}\left[\left(\Theta^{\dagger}-\frac{i\bm{\alpha}\cdot\bm{\tau}}{2}\Theta^{\dagger}\right)\left(\overleftarrow{\partial}_{\mu}-\frac{ia_{\mu}^{j}\sigma^{i}}{2}\right)\left(\vec{\partial}_{\mu}+\frac{ia_{\mu}^{j}\sigma^{i}}{2}\right)\left(\Theta+\Theta\frac{i\bm{\alpha}\cdot\bm{\tau}}{2}\right)\right]\\
 & =\mathcal{L}_{\text{kin}}^{\Theta}+\frac{1}{2}\text{tr}\left[\left(-\frac{i\bm{\alpha}\cdot\bm{\tau}}{2}\Theta^{\dagger}\right)\left(\overleftarrow{\partial}_{\mu}-\frac{ia_{\mu}^{j}\sigma^{i}}{2}\right)\left(\vec{\partial}_{\mu}+\frac{ia_{\mu}^{j}\sigma^{i}}{2}\right)\Theta\right]\\
 & \qquad+\frac{1}{2}\text{tr}\left[\Theta^{\dagger}\left(\overleftarrow{\partial}_{\mu}-\frac{ia_{\mu}^{j}\sigma^{i}}{2}\right)\left(\vec{\partial}_{\mu}+\frac{ia_{\mu}^{j}\sigma^{i}}{2}\right)\left(\Theta\frac{i\bm{\alpha}\cdot\bm{\tau}}{2}\right)\right].
\end{align}
When $\alpha$ is constant, then $(\mathcal{L}_{\text{kin}}^{\Theta})'=\mathcal{L}_{\text{kin}}^{\Theta}$
to all orders. When it varies in spacetime, on the other hand 
\begin{align}
(\mathcal{L}_{\text{kin}}^{\Theta})'-\mathcal{L}_{\text{kin}}^{\Theta} & =\frac{1}{2}\text{tr}\left[\left(-\frac{i\bm{\alpha}\cdot\bm{\tau}}{2}\Theta^{\dagger}\right)\left(\overleftarrow{\partial}_{\mu}-\frac{ia_{\mu}^{j}\sigma^{i}}{2}\right)\left(\vec{\partial}_{\mu}+\frac{ia_{\mu}^{j}\sigma^{i}}{2}\right)\Theta\right]\\
 & \qquad+\frac{1}{2}\text{tr}\left[\Theta^{\dagger}\left(\overleftarrow{\partial}_{\mu}-\frac{ia_{\mu}^{j}\sigma^{i}}{2}\right)\left(\vec{\partial}_{\mu}+\frac{ia_{\mu}^{j}\sigma^{i}}{2}\right)\left(\Theta\frac{i\bm{\alpha}\cdot\bm{\tau}}{2}\right)\right].
\end{align}
After expansion and simplification, the two terms above reduce to:
\begin{align}
\frac{1}{2}\text{tr}\left[\Theta^{\dagger}\left(\overleftarrow{\partial}_{\mu}-\frac{ia_{\mu}^{j}\sigma^{i}}{2}\right)\left(\vec{\partial}_{\mu}+\frac{ia_{\mu}^{j}\sigma^{i}}{2}\right)\left(\Theta\frac{i\bm{\alpha}\cdot\bm{\tau}}{2}\right)\right]
 & =\frac{1}{2}\text{tr}\left[\Theta^{\dagger}\left(\overleftarrow{\partial}_{\mu}-\frac{ia_{\mu}^{j}\sigma^{i}}{2}\right)\Theta\frac{i\partial_{\mu}\bm{\alpha}\cdot\bm{\tau}}{2}\right]\\
 & \qquad+\frac{1}{2}\text{tr}\left[\Theta^{\dagger}\left(\overleftarrow{\partial}_{\mu}-\frac{ia_{\mu}^{j}\sigma^{i}}{2}\right)\left(\vec{\partial}_{\mu}+\frac{ia_{\mu}^{j}\sigma^{i}}{2}\right)\Theta\times\frac{i\bm{\alpha}\cdot\bm{\tau}}{2}\right],
\end{align}
and likewise
\begin{align}
\frac{1}{2}\text{tr}\left[\left(-\frac{i\bm{\alpha}\cdot\bm{\tau}}{2}\Theta^{\dagger}\right)\left(\overleftarrow{\partial}_{\mu}-\frac{ia_{\mu}^{j}\sigma^{i}}{2}\right)\left(\vec{\partial}_{\mu}+\frac{ia_{\mu}^{j}\sigma^{i}}{2}\right)\Theta\right] &= -\frac{1}{2}\text{tr}\left[\frac{i\partial_{\mu}\bm{\alpha}\cdot\bm{\tau}}{2}\Theta^{\dagger}\left(\vec{\partial}_{\mu}+\frac{ia_{\mu}^{j}\sigma^{i}}{2}\right)\Theta\right]\\
 & \qquad-\frac{1}{2}\text{tr}\left[\frac{i\bm{\alpha}\cdot\bm{\tau}}{2}\times\Theta^{\dagger}\left(\overleftarrow{\partial}_{\mu}-\frac{ia_{\mu}^{j}\sigma^{i}}{2}\right)\left(\vec{\partial}_{\mu}+\frac{ia_{\mu}^{j}\sigma^{i}}{2}\right)\Theta\right].
\end{align}
By cyclicity of the trace, what remains after cancellation is:
\begin{align}
(\mathcal{L}_{\text{kin}}^{\Theta})'-\mathcal{L}_{\text{kin}}^{\Theta}=\partial_{\mu}\alpha^{j}\cdot\frac{1}{2}\text{tr}\left[\Theta^{\dagger}\left(\overleftarrow{\partial}_{\mu}-\frac{ia_{\mu}^{j}\sigma^{i}}{2}\right)\Theta\frac{i\tau^{j}}{2}-\frac{i\tau^{j}}{2}\Theta^{\dagger}\left(\vec{\partial}_{\mu}+\frac{ia_{\mu}^{j}\sigma^{i}}{2}\right)\Theta\right],
\end{align}
so that the conserved current is
\begin{align}
\bm{\mathcal{J}}_{\mu}\propto\text{tr}\left[\frac{\bm{\tau}}{2}\frac{\Theta^{\dagger}\left(\vec{\partial}_{\mu}+\frac{ia_{\mu}^{j}\sigma^{i}}{2}\right)\Theta-\Theta^{\dagger}\left(\overleftarrow{\partial}_{\mu}-\frac{ia_{\mu}^{j}\sigma^{i}}{2}\right)\Theta}{2i}\right]=\text{tr}\left[\frac{\bm{\tau}}{2}\frac{\Theta^{\dagger}\vec{D}_{\mu}\Theta-\Theta^{\dagger}\overleftarrow{D}_{\mu}\Theta}{2i}\right].
\end{align}
Because of the pseudospin Pauli vector $\bm{\tau}/2$, this bosonic object transforms as a vector under global pseudospin rotations $\Theta\to\Theta e^{i\bm{\theta}\cdot\bm{\tau}/2}.$ Indeed, its $\tau^z$ component is like the density of low-energy Cooper pairs, while $\tau^{\pm}$ serve as Cooper pair annihilation/creation operators. Moreover, it is gauge-invariant since $D_{\mu}\Theta$ is gauge-covariant. Finally, since it is the conserved current, then at the critical point all components will have protected scaling dimension 2. Altogether, this SU(2) formulation confirms the physical conclusions made in Ref.~\cite{Lee2018}, by working here in a formalism manifestly invariant under the pseudospin symmetry.

\section{Electron Pairing from Small-$U$ Diagrammatic Expansion} \label{app_SemiAnalytic}

\subsection{Band basis in the doubly-folded Brillouin zone}

Let's work in the Imaginary C6 gauge, which has four sublattice
sites $\mathcal{S}=\{A,B,C,D\},$ arranged as in Fig.~\ref{fig:two_gauges}(a). While $T_{1}$
and $T_{2}$ anti-commute in the odd-fermion-parity sector, $(T_{1})^{2}$
and $(T_{2})^{2}$ commute. Thus, given displacements $\bm{d}\in\bar{\mathcal{L}}\equiv2\mathcal{L}$
(\emph{i.e.} in the $A$ sublattice), then the operators $T_{\bm{d}}$
have unambiguous action and \emph{form a group.} In fact, in the Imaginary
$C_{6}$ gauge with single-site translations defined as in Eq.~\eqref{eq:T1_T2},
these two-site translation operators are \emph{pure coordinate shift},
as shown in Eq.~\eqref{eq:twosite_pure_shift} above. Therefore we can readily apply Bloch's
theorem with a four-site unit cell.
Let $\overline{\mathrm{BZ}}$ denote the ``doubly-folded'' Brillouin zone corresponding to $\bar{\mcL}$.
The Fourier-transformed
electron operators are
\begin{align} \label{eq:electron_fourier}
c_{\bm{k},\gamma}^{\dagger}=\frac{1}{\sqrt{|\bar{\mathcal{L}}|}}\sum_{\bm{d}\in\bar{\mathcal{L}}}e^{i\bm{k}\cdot(\bm{d}+\bm{\gamma})}T_{\bm{d}}c_{\bm{0}+\bm{\gamma}}^{\dagger}T_{\bm{d}}^{\dagger}=\frac{1}{\sqrt{|\bar{\mathcal{L}}|}}\sum_{\bm{d}\in\bar{\mathcal{L}}}e^{i\bm{k}\cdot(\bm{d}+\bm{\gamma})}c_{\bm{d}+\bm{\gamma}}^{\dagger},
\end{align}
where we used the fact that $T_{\bm{d}}$ is pure coordinate shift
for $\bm{d}\in\bar{\mathcal{L}}.$

Now let's Fourier transform the hopping Hamiltonian, namely
\begin{align}
h=-\sum_{\bm{r},\bm{r}'}t_{\bm{r}\bm{r}'}c_{\bm{r}}^{\dagger}c_{\bm{r}'}=-\sum_{\gamma,\gamma'}\sum_{\bm{d},\bm{d}'\in\bar{\mathcal{L}}}t_{\bm{d}+\bm{\gamma},\bm{d}'+\bm{\gamma}'}c_{\bm{d}+\bm{\gamma}}^{\dagger}c_{\bm{d}'+\bm{\gamma}'},
\end{align}
where $\gamma,\gamma'$ run over the sublattices $\mathcal{S}$ and
\begin{align}
t_{\bm{d}+\bm{\gamma},\bm{d}'+\bm{\gamma}'}=t_{\gamma,\gamma'}(\bm{d}-\bm{d}')
\end{align}
is manifestly translation-invariant on the scale of $2\mathcal{L}$. Inverting the electron Fourier transform
Eq.~\eqref{eq:electron_fourier}, we obtain
\begin{align}
c_{\bm{d}+\bm{\gamma}}^{\dagger}=\sum_{\bm{k}\in\overline{\mathrm{BZ}}}e^{-i\bm{k}\cdot(\bm{d}+\bm{\gamma})}c_{\bm{k},\gamma}^{\dagger},
\end{align}
and substituting it in gives
\begin{align}
h & =-\frac{1}{|\bar{\mathcal{L}}|}\sum_{\gamma,\gamma'}\sum_{\bm{d},\bm{d}'\in\bar{\mathcal{L}}}t_{\bm{d}+\bm{\gamma},\bm{d}'+\bm{\gamma}'}\left(\sum_{\bm{k}\in\overline{\mathrm{BZ}}}e^{-i\bm{k}\cdot(\bm{d}+\bm{\gamma})}c_{\bm{k},\gamma}^{\dagger}\right)\left(\sum_{\bm{k}'\in\overline{\mathrm{BZ}}}e^{i\bm{k}'\cdot(\bm{d}'+\bm{\gamma}')}c_{\bm{k}',\gamma'}\right)\\
 & =\sum_{\bm{k},\bm{k}'\in\overline{\mathrm{BZ}}}\sum_{\gamma,\gamma'}\left[  \sum_{\bm{\delta}\in\bar{\mathcal{L}}} -e^{-i\bm{k}\cdot\bm{\delta}} e^{-i\bm{k}\cdot\bm{\gamma}}t_{\bm{\gamma},\bm{\gamma}'}(\bm{\delta}) e^{i\bm{k}'\cdot\bm{\gamma}'}\right]\left(\frac{1}{|\bar{\mathcal{L}}|}\sum_{\bm{d}'\in\bar{\mathcal{L}}}e^{i(\bm{k}'-\bm{k})\cdot\bm{d}'}\right)c_{\bm{k},\gamma}^{\dagger}c_{\bm{k}',\gamma'}\\
 & =\sum_{\bm{k}\in\overline{\mathrm{BZ}}}\sum_{\gamma,\gamma'}c_{\bm{k},\gamma}^{\dagger}h_{\gamma,\gamma'}(\bm{k})c_{\bm{k},\gamma'},
\end{align}
where
\begin{align}
h_{\gamma,\gamma'}(\bm{k})= e^{-i\bm{k}\cdot\bm{\gamma}} \left(-\sum_{\bm{\delta}\in\bar{\mathcal{L}}} e^{-i\bm{k}\cdot\bm{\delta}} t_{\bm{\gamma},\bm{\gamma}'}(\bm{\delta}) \right)e^{i\bm{k}'\cdot\bm{\gamma}'}.
\end{align}
Using $t_{\gamma',\gamma}(-\bm{\delta})^{*}=t_{\gamma,\gamma'}(\bm{\delta}),$
one can show that $h_{\gamma,\gamma'}(\bm{k})$ is a Hermitian matrix
in the sublattice. We plot the higher of the two resulting bands in
Fig.~1(c) of the main text. Note that the lower and upper bands are each two-fold degenerate, a consequence of our having folded the lattice-scale Brillouin zone
not once but twice (in order to accommodate the $2\times2$-site unit
cell and $C_{6}$ rotations).

Let us denote the magnetic reciprocal lattice vectors of the magnetic
lattice by
\begin{align}
\bar{\mathcal{L}}^{-1}\in\mathrm{span}\left\{ \bm{b}_{1}/2,\bm{b}_{2}/2\right\} 
\end{align}
(where the ``span'' is over integer linear combinations), defined
equivalently as those vectors for which $\bm{\delta}\in\bar{\mathcal{L}}\implies\bm{G}\cdot\bm{\delta}\in2\pi\mathbb{Z}.$
Under shifts by such vectors, the Hamiltonian transforms as
\begin{align}
h_{\gamma,\gamma'}(\bm{k}+\bm{G})=e^{-i\bm{G}\cdot\bm{\gamma}}h_{\gamma,\gamma'}(\bm{k})e^{i\bm{G}\cdot\bm{\gamma}'}.
\end{align}
As $4\times4$ matrices in the sublattice degree of freedom, $h(\bm{k}+\bm{G})$
is therefore unitarily-related to $h(\bm{k})$ and has the same spectrum.
We obtain eigenstates by diagonalization:
\begin{align}
h(\bm{k})u(\bm{k})=u(\bm{k})E(\bm{k}),
\end{align}
where $E(\bm{k})$ is a diagonal matrix of energies, while $u_{\gamma,n}(\bm{k})$
is a unitary matrix whose columns are labeled by band $n$ and whose
rows are labeled by sublattice $\gamma.$ Correspondingly, the Bloch
fermion creation operators are
\begin{align}
f_{n}^{\dagger}(\bm{k})=\sum_{\gamma}c_{\bm{k},\gamma}^{\dagger}u_{\gamma,n}(\bm{k}),
\end{align}
which is easily shown to diagonalize the hopping Hamiltonian and satisfy
the usual fermion anti-commutation relations.

\subsection{Perturbative demonstration of odd-angular momentum spin-singlet pairing}

We rewrite the model in momentum space 
\begin{equation} \label{eq:GenericRGHamiltonianStart} \begin{split}
\mathcal{H} & = \sum_{ 12 } c_{2}^\dagger h_{2,1} c_{1} + \frac{1}{2N}  \sum_{ 1234 } c_{4}^\dagger c_{3}^\dagger \Gamma_{43,21} c_{2} c_{1} ,
\end{split} \end{equation}
with $N$ the number of unit cells, and $i = ({k}_i, o_i, \sigma_i)$ a generalized index gathering the single particle momentum ${k}_i$ belonging to the Brillouin zone (BZ), the sublattice index $o_i$, and the spin $\sigma_i$. Due to momentum and spin conservation, and because our model only features a Hubbard interaction, the parameters in this Hamiltonian can be simplified as 
\begin{equation} \label{eq:SimplerFormInteraction} \begin{split}
h_{2,1} & = \delta_{k_1, k_2} \delta_{\sigma_1, \sigma_2} h_{o_2, o_1} (k_1) , \\ \Gamma_{43,21} & = U \delta_{s_4,s_1, \uparrow} \delta_{s_3,s_2, \downarrow} \delta_{o_4, o_3, o_2, o_1} \delta_{k_1+k_2,k_3+k_4} ,
\end{split} \end{equation}
Renormalization of the two-body scattering vertex is described, to second order in many-body perturbation theory, by the five diagrams~\cite{KohnLuttinger1965}:
\begin{equation} \label{eq:FiveKLDiagramsapp}
\begin{tikzpicture}[baseline=0cm,scale=0.8]
\draw (-0.75,0.5) -- (0.75,0.5); \draw (-0.75,-0.5) -- (0.75,-0.5); 
\draw[snake it] (-0.33,0.5) -- (-0.33,-0.5); \draw[snake it] (0.33,0.5) -- (0.33,-0.5);
\end{tikzpicture} , \, 
\begin{tikzpicture}[baseline=0cm,scale=0.8]
\draw (-0.75,0.5) -- (0.75,0.5); \draw (-0.75,-0.5) -- (0.75,-0.5); 
\draw[snake it] (-0.33,0.5) -- (0.33,-0.5); \draw[snake it] (0.33,0.5) -- (-0.33,-0.5);
\end{tikzpicture} , \, 
\begin{tikzpicture}[baseline=0cm,scale=0.8]
\draw (-0.75,-0.5) -- (0.75,-0.5); \draw (-0.75,0.5) -- (-0.33,0.5) -- (0,0) -- (0.33,0.5) -- (0.75,0.5); 
\draw[snake it] (0,-0.5) -- (0,0); \draw[snake it] (0.33,0.5) -- (-0.33,0.5);
\end{tikzpicture} , \, 
\begin{tikzpicture}[baseline=0cm,scale=0.8]
\draw (-0.75,0.5) -- (0.75,0.5); \draw (-0.75,-0.5) -- (-0.33,-0.5) -- (0,0) -- (0.33,-0.5) -- (0.75,-0.5); 
\draw[snake it] (0,0.5) -- (0,0); \draw[snake it] (0.33,-0.5) -- (-0.33,-0.5);
\end{tikzpicture} , \, 
\begin{tikzpicture}[baseline=0cm,scale=0.8]
\draw (0,0) circle (0.25); 
\draw (-0.75,0.5) -- (0.75,0.5); \draw (-0.75,-0.5) -- (0.75,-0.5); 
\draw[snake it] (0.,0.5) -- (0.,0.25); \draw[snake it] (0.,-0.5) -- (0,-0.25);
\end{tikzpicture} , \, 
\end{equation}
where curvy and straight lines respectively denote interaction events and single-particle propagators. The evaluation of these diagrams is detailed in App.~D of Ref.~\cite{crepel2022unconventional}, producing a renormalized interaction vertex $\Gamma^{\rm eff} = \Gamma + \delta \Gamma$, with
\begin{equation} \label{eq:AllKLTerms}
\delta \Gamma_{43,21} = - \frac{1}{N} \sum_{abcd} \chi_{dc,ba}^- \Gamma_{43,ba} \Gamma_{dc,21} + \chi_{dc,ba}^+ [\Gamma_{4c,2a} \Gamma_{d3,b1} + \Gamma_{c3,2a} \Gamma_{d4,b1} + \Gamma_{4c,a1} \Gamma_{d3,b2} - N_f \Gamma_{4c,a1} \Gamma_{d3,2b}] , 
\end{equation}
where the terms are ordered as their corresponding diagrams in Eq.~\eqref{eq:FiveKLDiagramsapp} where $N_f = 2$ is the number of spin species. The sum over the generalized indices $(a,b,c,d)$ runs over momenta, orbital and spin indices. The single-particle propagators are characterized by the Bloch eigenvectors $\Psi_{k_i,n}^{o_i}$ with $n$ a band index, which are the same for both spin species. Finally, we have respectively denoted as $\chi^-$ and $\chi^+$ the particle-particle and particle-hole susceptibilities, whose explicit form is
\begin{align}\label{eq:SusceptibilitiesKL}
& \chi_{dc,ba}^{\epsilon = \pm} = \delta_{{k}_a}^{{k}_d} \delta_{{k}_b}^{{k}_c} \sum_{n,n'=\pm} \Psi_{k_d,n}^{o_d \, *} \Psi_{k_c,n'}^{o_c \, *} \Psi_{k_b,n'}^{o_b} \Psi_{k_a,n}^{o_a} \frac{f_\beta (\epsilon \xi_{{k}_a,n}) - f_\beta(\xi_{{k}_b,n'})}{\xi_{{k}_a,n} - \epsilon \xi_{{k}_b,n'}} , \notag
\end{align}
where $f_\beta(x) = 1/[1+e^{\beta x}]$ is the Fermi-Dirac distribution, and $\xi_i = \varepsilon_i-\mu$ measures energies with respect to the chemical potential $\mu$.

We are interested in the correction to the Cooper interaction $\delta \Gamma_{(k',\uparrow,o_4)(-k',\downarrow,o_3),(-k,\downarrow,o_2)(k,\uparrow,o_1)}$ at zero temperature when the chemical potential is in the middle of the gap. Several simplifications arise:
\begin{itemize}
\item First, the contributions of bubble and of the upper and lower wine glass diagrams vanish. This can be straightforwardly checked using the spin constraints on the bare vertex $\Gamma$.
\item Second, we can see that $\chi^-$ only involves intra-band processes that are most relevant in presence of a Fermi surface, \textit{i.e.}, at finite doping which is not the focus of this pairing calculation. This simply comes from $f_\beta(\xi<0)=1$ and $f_\beta(\xi>0)=0$.
\item This only leaves the crossed diagram.
\end{itemize}

Focusing on the Cooper channel, we evaluate this crossed diagram and project onto the given bands (in the Imaginary C6 gauge, there are two bands per valley above half-filling) to get
\begin{equation}
\delta \Gamma_{(k',\uparrow,n_4)(-k',\downarrow,n_3),(-k,\downarrow,n_2)(k,\uparrow,n_1)} = - U^2 \sum_{o,o'} \left( \Psi_{k',o}^{n_4} \Psi_{-k',o'}^{n_3} \right)^* \tilde\chi_{o',o}  (k'+k) \left( \Psi_{-k,o}^{n_2} \Psi_{k,o'}^{n_1} \right) , 
\end{equation}
where the orbital susceptibilities are defined by 
\begin{equation}
\tilde\chi_{o',o}  (q) = \frac{1}{N} \sum_p \chi_{(p+q,o')(p,o),(p,o')(p+q,o)}^+ . 
\end{equation}
Note that in spite of the overall sign in front of the sum, this interaction is mostly repulsive because the susceptibilities defined above are negative in the density-density channel where $o=o'$. 

If we focus on the effective scattering coefficient between electrons occupying the lowest single-particle states of the model, then we can replace $k$ and $k'$ by $\kappa/\kappa'$. Then there are only three orbital susceptibilities to compute $\tilde\chi(0, \kappa, \kappa')$  and four Bloch states to obtain in order to evaluate all the different Cooper scattering coefficients. This forms a matrix that we can diagonalize to identify the most attractive (if any). We have computed all of those terms, and have found a single attractive channel that we write in the basis where band \textbf{0} has $C_3$ eigenvalues $\omega$ at $\kappa/\kappa'$, bands \textbf{1} and \textbf{3} have zero angular momentum at those points, and band \textbf{2} has $C_3$ eigenvalue of $\omega^*$. All of these single-electron angular momenta were computed using the conventions for gauge and $C_6$ described in App.~\ref{app:imag_C6_gauge} above.

We focus on the conduction bands \textbf{2} and \textbf{3}. The pairing channel we find is a spin-singlet and carries $C_3$ angular momentum (one electron of the pair in each band) in our choice of gauge and for our choice of rotation symmetry $C_3$.
Examining at the pair closely, it is antisymmetric in the band index and in the valley index (hence fully antisymmetric under particle exchange, as required).
Altogether, we can write the pair down as:
\begin{align}
\hat{\Delta} = c_{\kappa',\uparrow,\omega^*} c_{\kappa,\downarrow,1} + c_{\kappa,\uparrow,1} c_{\kappa',\downarrow,\omega^*} - c_{\kappa',\uparrow,1} c_{\kappa,\downarrow,\omega^*} - c_{\kappa,\uparrow,\omega^*} c_{\kappa',\downarrow,1} .
\end{align}
The valley antisymmetry, corresponding to eigenvalue $-1$ under $C_2$, makes explicit the \textit{odd site-centered angular momentum nature of the pair}.
Recall the discussion in the main text that $C_3\to (e^{2\pi i/3})^{N_F} C_3$ may be redefined, thus modifying the $C_3$ eigenvalue, while preserving the identity $(C_3)^3=1$.
Therefore, to prevent confusion, in the main text we have replaced the $C_3$ eigenvalue labels $1$ and $\omega^*$ with an index $n\in \{1,2\}$ which likewise labels the degenerate bands at each of $\kappa$ and $\kappa'$.

\end{document}